\documentclass[3p,number,times,fleqn]{elsarticle}

\makeatletter
\def\ps@pprintTitle{%
  \let\@oddhead\@empty
  \let\@evenhead\@empty
  \let\@evenfoot\@oddfoot
}
\makeatother

\usepackage{graphicx}
\usepackage{color}
\usepackage{mathtools}
\usepackage{amsmath}
\usepackage{amssymb}
\usepackage[only,llbracket,rrbracket]{stmaryrd}
\usepackage[misc]{ifsym}
\usepackage{bm}                    
\usepackage{setspace}
\usepackage{xspace}
\usepackage{hyphenat}
\usepackage{comment}
\usepackage{array} 
\usepackage{float} 
\usepackage{multirow}
\floatstyle{ruled}
\newfloat{Fbox}{thp}{lop}
\floatname{Fbox}{Box}
\usepackage[flushmargin]{footmisc} 
\usepackage[colorlinks,pdfpagelabels,bookmarksopen,bookmarksnumbered,citecolor=blue,urlcolor=blue,linkcolor=blue]{hyperref}

\newdefinition{rmk}{Remark}
\newproof{pf}{Proof}
\newproof{pot}{Proof of Theorem \ref{thm2}}

\newcommand{\avg}[1]{\langle#1\rangle}
\newcommand{\bs}[1]{\mathbf{#1}}
\newcommand{\bms}[1]{\bm{#1}}
\newcommand{\bigA}{\mathop{\vphantom{\sum}\mathchoice{\vcenter{\hbox{\huge A}}}{\vcenter{\hbox{\Large A}}}{\mathrm{A}}{\mathrm{A}}}\displaylimits} 

\newcommand{\new}[1]{{\color{black}#1}}
\newcommand{\old}[1]{\iffalse#1\fi}

\begin{document}

\title{Computational design of locally resonant acoustic metamaterials}

\author[cimne,eseiaat]{D. Roca}
\author[cimne,eseiaat]{D. Yago}
\author[cimne,eseiaat]{J. Cante}
\author[cimne,etseccpb]{O. Lloberas-Valls}
\author[cimne,etseccpb]{J. Oliver\corref{cor}}
\ead{oliver@cimne.upc.edu}

\cortext[cor]{Corresponding author}

\address[cimne]{
	Centre Internacional de M\`{e}todes Num\`{e}rics en Enginyeria (CIMNE)\\ 
	Campus Nord UPC, M\`{o}dul C-1 101, c/ Jordi Girona 1-3, 08034 Barcelona, Spain
}
\address[eseiaat]{
	Escola Superior d'Enginyeries Industrial, Aeroespacial i Audiovisual de Terrassa (ESEIAAT)\\
	Technical University of Catalonia (Barcelona Tech), Campus Terrassa UPC, c/ Colom 11, 08222 Terrassa, Spain
}
\address[etseccpb]{
	E.T.S d'Enginyers de Camins, Canals i Ports de Barcelona (ETSECCPB)\\
	Technical University of Catalonia (Barcelona Tech), Campus Nord UPC, M\`{o}dul C-1, c/ Jordi Girona 1-3, 08034 Barcelona, Spain
}

\begin{abstract}
	The so-called Locally Resonant Acoustic Metamaterials (LRAM) are considered for the design of specifically engineered devices capable of stopping waves from propagating in certain frequency regions (bandgaps), this making them applicable for acoustic insulation purposes. This fact has inspired the design of a new kind of lightweight acoustic insulation panels with the ability to attenuate noise sources in the low frequency range (below 5000 Hz) without requiring thick pieces of very dense materials. A design procedure based on different computational mechanics tools, namely, (1) a multiscale homogenization framework, (2) model order reduction strategies and (3) topological optimization procedures, is proposed. It aims at attenuating sound waves through the panel for a target set of resonance frequencies as well as maximizing the associated bandgaps. The resulting design's performance is later studied by introducing viscoelastic properties in the coating phase, in order to both analyse their effects on the overall design and account for more realistic behaviour. The study displays the emerging field of Computational Material Design (CMD) as a computational mechanics area with enormous potential for the design of metamaterial-based industrial acoustic parts. 
\end{abstract}


\maketitle

\section{Motivation}
\label{sec_motivation}
The concept of Locally Resonant Acoustic Metamaterials (LRAM) has been object of growing interest among the scientific and technological communities in recent years. The notion of \textit{metamaterials} emerged in the late 1990s as a new kind of engineered materials, by optimizing their morphology or arrangement at lower scales, capable of exhibiting properties ``on demand'' that are not found in naturally occurring materials. In the specific case of acoustic metamaterials, these properties involve exotic or counter-intuitive behaviour caused by the interaction of the material with acoustic or elastic waves' propagation features. These type of materials could be used, for instance, to design acoustic insulation panels that target specific frequency bands (especially in the low-frequency range, i.e. below 5000 Hz, where most sources of environmental noise are produced). In contrast to conventional acoustic panels, which require either a large thickness or high mass density in order to provide effective sound attenuation at lower frequencies, LRAM-based panels can achieve good levels of transmission loss in the whole frequency range of interest with relatively thin, lightweight designs.  

Scientific research in the field of metamaterials started in the late 19th century with the works of Floquet and Rayleigh among others, who studied phenomena related to the propagation of waves in periodic systems. However, it was not until the beginning of this century when the first implementation of an acoustic metamaterial capable of stopping waves from transmitting in a certain frequency band was reported by \citet{Liu2000}. Later on, \citet{Ho2005} and \citet{Calius2009} performed other experimental demonstrations with silicone rubber-coated metal spheres embedded in polymer matrices, while more recently, \citet{Claeys2016} have carried out tests with a fully 3D-printed design with internal resonators capable of achieving also good attenuation properties in the low-frequency range. The idea of LRAM-based insulation panels has already been explored, both theoretically and experimentally, for membrane-type  \citep{Yang2010,Zhang2013} and plate-type \citep{Badreddine2012} acoustic metamaterials. 

On the other hand, computational models for the study and characterization of LRAMs have been traditionally focused on periodically repeated microstructures where Bloch-Floquet boundary conditions can be applied \citep{Hussein2009,Hussein2014}. More recent developments include the works of \citet{Sridhar2016,Sridhar2017} and \citet{Roca2018}, in which computational homogenization frameworks accounting for inertial effects have been proposed, capable of capturing the local resonance phenomena that characterizes such kind of materials. 

Aiming at trying to enlarge the effective attenuation band achieved by the local resonance phenomenon in metamaterials is a challenging task, since the so-called frequency bandgaps produced by typical LRAM designs, which are the source of their attenuation capabilities, tend to be too narrow in the frequency spectrum. In this regard, several proposals have been made recently in the literature in order to find a solution for this problem. For instance, \citet{Matsuki2014} proposed a topology optimization-based method to obtain optimal LRAM configurations with multiple attenuation peaks, which can be considered one of the first attempts to apply optimization procedures to LRAM materials. Other approaches are focused on taking advantage of the viscoelastic nature of the coating materials in typical LRAM configurations. The first notions on the beneficial effects of viscous damping in acoustic metamaterials were reported by \citet{Hussein2013}, who introduced the concept of \textit{metadamping} to refer to the damping emergence phenomenon produced as a result of combining the effects of local resonance with viscous dissipation. \new{The concept of metadamping has also been explored in more detail in subsequent works \cite{Frazier2015b} and the idea of acoustic metamaterial configurations based on this (phononic resonators) has been proposed by \citet{depauw2018}}. This phenomenon has also been studied more recently in the context of acoustic metamaterials in the work of \citet{Manimala2014}, where they show, by means of an analytical approach, the dispersion properties of LRAMs with different models of viscoelastic (dissipative) behaviour for the coating material, and the works of \citet{Krushynska2016} and \citet{Lewinska2017}, in which generalized viscoelastic modelling is introduced in the study of the attenuation performance of LRAMs. 

In this paper, the authors attempt at setting a computational based methodology for the modelling, analysis and design of metamaterial parts exhibiting local resonance phenomena by combining three well-established complementary computational tools: (1) a multiscale hierarchical homogenization procedure specifically devised for acoustic problems, described in \citet{Roca2018}, (2) the exploitation of Reduced Order Modelling (ROM) techniques, to minimize the resulting computational cost, based on selective projections of the RVE behaviour onto the space spanned by its natural modes and (3) topology optimization techniques. 

A design strategy is proposed to optimize the attenuation performance of an insulation panel made of LRAMs for a target band of frequencies, by optimizing the topology of the material at the mesoscale. This allows considering its industrial manufacture by means of emergent 3D-printing or similar techniques. The proposed design strategy is based on (a) fitting the lower bound of the target band with some natural resonance frequency of the material at the design scale, and (b) maximizing the target band's bandwidth. The resulting design exhibits acoustic insulation properties much improved, in comparison to those that could have been obtained by simple trial-and-error procedures, which proves the benefits of the considered CMD methodology. 

\section{Multiscale modelling of LRAMs}
\label{sec_LRAM_homog}
The computational homogenization framework introduced in \citet{Roca2018} has been used here as the base model upon which the methodology for the design of LRAMs will be built. The framework can be applied to problems where a separation of scales is present, for instance, allowing us to identify a representative volume element (RVE), typically a unit cell, in a macroscopic structure. This separation of scales is established in terms of the macroscopic characteristic wavelength, $\lambda$, which has to be larger than the characteristic size of the lower scale, $\ell_\mu$, otherwise the validity of the homogenization model cannot be guaranteed. This is not an impediment to deal with LRAMs considering they are designed to operate in the low-frequency regime where the condition
\begin{equation}
	\label{eq_separation_scales_condition}
	\lambda \gg \ell_\mu
\end{equation}
is easily satisfied. In fact, expression \eqref{eq_separation_scales_condition} is also a condition required for local resonance phenomena to arise  \citep{Hussein2014}. 

For clarity purposes, from now on magnitudes referring to the microscale will be subscripted by $\mu$, in order to distinguish them from their macroscopic counterparts. Additionally, space coordinates for the macroscale will be referred by $\bs{x}$, while those for the microscale will be referred by $\bs{y}$, when necessary.

The framework is grounded on a generalization accounting for inertial effects of the classical Hill-Mandel principle \citep{Blanco2016}, in which the macroscale is assumed to behave as a Cauchy's continua, thus satisfying the classical postulates of linear and angular momentum:
\begin{align}
&\nabla_\bs{x} \cdot \bms{\sigma}(\bs{x},t) = \dot{\bs{p}}(\bs{x},t), \label{eq_macro_linear_momentum}\\
&\bms{\sigma}(\bs{x},t) = \bms{\sigma}^\text{T}(\bs{x},t), \quad \forall \bs{x} \in \Omega, \quad \forall t \in [0,T],
\end{align}
where $\bms{\sigma}$ is the macroscopic stress, $\dot{\bs{p}}$ is the macroscopic inertial force, $t$ refers to the time variable, $\dot{(\bullet)}$ stands for the time derivative of $(\bullet)$, and $\Omega$ refers to the spatial macroscopic domain. Note that body forces have not been considered, for the sake of simplicity, since they are not relevant in the context of acoustic problems that are tackled here.

By applying an energetic equivalence between scales, which is given by a variational statement of the generalized Hill-Mandel principle, \new{ i.e.
\begin{equation}
	\dot{\bs{p}}\cdot\dot{\bs{u}} + \bms{\sigma} : \dot{\bms{\varepsilon}} = \avg{\dot{\bs{p}}_\mu \cdot \dot{\bs{u}}_\mu + \bms{\sigma}_\mu : \nabla^\text{S}_\bs{y}\dot{\bs{u}}_\mu}_{\Omega_\mu}, \quad \forall \dot{\bs{u}}, \dot{\bms{\varepsilon}} \quad \text{and} \quad \forall \dot{\bs{u}}_\mu \in \bms{\mathcal{U}}_\mu;
\end{equation}}
along with kinematic restrictions that link the macroscopic displacements, $\bs{u}$, and strains, $\bms{\varepsilon} = \nabla_\bs{x}^\text{S}\bs{u}$, with their microscale counterparts, $\bs{u}_\mu$ and $\bms{\varepsilon}_\mu = \nabla_\bs{y}^\text{S}\bs{u}_\mu$, \new{namely
\begin{equation}
\label{eq_min_restric}
	\bms{\mathcal{U}}_\mu := \{\bs{u}_\mu : \Omega_\mu \times [0,T] \rightarrow \mathbb{R}^m \ | \ \avg{\bs{u}_\mu}_{\Omega_\mu} = \bs{u}; \ \avg{\nabla^\text{S}_\bs{y}\bs{u}_\mu}_{\Omega_\mu} = \bms{\varepsilon} \},
\end{equation}}
one can obtain a Lagrange-extended dynamic system of equations for the RVE \new{in which the Lagrange multipliers, $\bms{\beta}$ and $\bms{\lambda}$ respectively, corresponding to the reactions to the minimal kinematic restrictions given by equation \eqref{eq_min_restric} can be identified as
\begin{align}
	\label{eq_beta}
	&\bms{\beta} = \avg{\dot{\bs{p}}_\mu}_{\Omega_\mu} = \dot{\bs{p}} \\
	\label{eq_lambda}
	&\bms{\lambda} = \avg{\bms{\sigma}_\mu + \dot{\bs{p}}_\mu \otimes^\text{S}(\bs{y}-\bs{y}^{(0)})}_{\Omega_\mu}.
\end{align}
An in-depth explanation of the theory and further details on the derivation of these terms can be found in \citet{Roca2018}. Note that the angle brackets symbol is used to refer to the average volume integral over the RVE, i.e. $\avg{(\bullet)}_{\Omega_\mu} = \int_{\Omega_\mu} (\bullet) \text{d}\Omega$.} 

After a Galerkin-based finite elements discretization, this so-called extended RVE system has the form:
\begin{equation}
\label{eq_micro_extended_system}
	\begin{bmatrix}
		\mathbb{M}_\mu & \bs{0} & \bs{0} \\
		\bs{0} & \bs{0} & \bs{0} \\
		\bs{0} & \bs{0} & \bs{0}
	\end{bmatrix}
	\begin{bmatrix}
		\ddot{\hat{\bs{u}}}_\mu \\
		\ddot{\bms{\beta}} \\
		\ddot{\bms{\lambda}}
	\end{bmatrix} + 
	\begin{bmatrix}
		\mathbb{K}_\mu & -\mathbb{N}_\mu^\text{T} & -\mathbb{B}_\mu^\text{T} \\
		-\mathbb{N}_\mu & \bs{0} & \bs{0} \\
		-\mathbb{B}_\mu & \bs{0} & \bs{0}
	\end{bmatrix}
	\begin{bmatrix}
		\hat{\bs{u}}_\mu \\
		\bms{\beta} \\
		\bms{\lambda}
	\end{bmatrix} = 
	\begin{bmatrix}
		\bs{0} \\
		-\bs{u} \\
		-\bms{\varepsilon}
	\end{bmatrix},
\end{equation}
where $\mathbb{M}_\mu$ and $\mathbb{K}_\mu$ are the standard mass and stiffness matrices of the RVE system, $\hat{\bs{u}}_\mu$ is the column vector with the nodal values for microscale displacement field, $\bs{u}_\mu(\bs{y},t)$, while $\bms{\beta}$ and $\bms{\lambda}$ are, respectively, the Lagrange multipliers associated to the kinematic restrictions over displacements and their gradient, which are imposed by the matrices $\mathbb{N}_\mu$ and $\mathbb{B}_\mu$, respectively.

Note that, in the system \eqref{eq_micro_extended_system}, the displacement and strain of the associated point, $\bs{u}(\bs{x},t)$, $\bms{\varepsilon}(\bs{x},t)$, become \textit{actions} and, \old{as reported in \citet{Roca2018}}\new{as indicated by equations \eqref{eq_beta} and \eqref{eq_lambda}}, one can relate the \textit{resulting} Lagrange multipliers, $\bms{\beta}(\bs{x},t)$ and $\bms{\lambda}(\bs{x},t)$, with the macroscopic inertial force and stress at that point, $\dot{\bs{p}}(\bs{x},t)$ and $\bms{\sigma}(\bs{x},t)$, respectively. As it is shown in \citet{Roca2018}, the system \eqref{eq_micro_extended_system} can be split into:
\begin{itemize}
	\item[1)] \textit{Quasi-static system} ($\bs{u}=\bs{0}$)
	\begin{equation}
	\label{eq_micro_quasi-static_system}
		\begin{bmatrix}
			\mathbb{K}_\mu & -\mathbb{B}_\mu^\text{T} \\
			-\mathbb{B}_\mu & \bs{0}
		\end{bmatrix}
		\begin{bmatrix}
			\hat{\bs{u}}_\mu^{(1)} \\
			\bms{\lambda}^{(1)}
		\end{bmatrix} =
		\begin{bmatrix}
			\bs{0} \\
			-\bms{\varepsilon}
		\end{bmatrix}.
	\end{equation}
	\item[2)] \textit{Inertial system} ($\bms{\varepsilon}=\bs{0}$)
	\begin{equation}
	\label{eq_micro_inertial_system}
		\begin{bmatrix}
		\mathbb{M}_\mu & \bs{0} \\
		\bs{0} & \bs{0}
		\end{bmatrix}
		\begin{bmatrix}
		\ddot{\hat{\bs{u}}}_\mu^{(2)} \\
		\ddot{\bms{\beta}}^{(2)}
		\end{bmatrix} +
		\begin{bmatrix}
			\mathbb{K}_\mu & -\mathbb{N}_\mu^\text{T} \\
			-\mathbb{N}_\mu & \bs{0}
		\end{bmatrix}
		\begin{bmatrix}
			\hat{\bs{u}}_\mu^{(2)} \\
			\bms{\beta}^{(2)}
		\end{bmatrix} =
		\begin{bmatrix}
			\bs{0} \\
			-\bs{u}
		\end{bmatrix}.
	\end{equation}
\end{itemize}
The split is based on considering each action separately in each subsystem along with certain hypotheses, which are suitable for the study of local resonance phenomena in acoustic problems. In particular:
\begin{itemize}
	\item[a)] The macroscopic strain accelerations are negligible, i.e. $\ddot{\bms{\varepsilon}} \approx \bs{0}$, allowing for the system \eqref{eq_micro_quasi-static_system} to actually behave quasi-statically, thus $\bms{\beta}^{(1)} = \dot{\bs{p}}^{(1)} \approx \bs{0}$.
	\item[b)] The RVE's topology exhibits symmetry with respect to its geometric centre, which allows us to assume, for the subsystem \eqref{eq_micro_inertial_system}, $\bms{\lambda}^{(2)} = \bms{\sigma}^{(2)} \approx \bs{0}$.
\end{itemize} 

From the quasi-static part of the system, one can derive an expression for the macroscopic stress that reads
\begin{equation}
	\label{eq_macro_stress}
	\bms{\sigma}(\bs{x},t) \approx \bs{C}^\text{eff}(\bs{x}) : \bms{\varepsilon}(\bs{x},t),
\end{equation}
where $\bs{C}^\text{eff}$ is an effective constitutive tensor. 

On the other hand, from the inertial subsystem, it is possible to obtain the macroscopic inertial force as
\begin{equation}
	\label{eq_macro_inertia}
	\dot{\bs{p}}(\bs{x},t) \approx \bar{\rho}(\bs{x})\ddot{\bs{u}}(\bs{x},t) + \mathbb{Q}(\bs{x}) \cdot \ddot{\bs{q}}_\mu(\bs{x},t)
\end{equation}
where $\bar{\rho}$ is the RVE average mass density and the second term in equation \eqref{eq_macro_inertia} represents the contribution of coupled micro-inertial effects (through the matrix $\mathbb{Q}$) whose behaviour is dictated by the reduced set of uncoupled equations resulting from the modal projection of the inertial subsystem \eqref{eq_micro_inertial_system}
\begin{equation}
\label{eq_micro_inertial_system_reduced}
	\ddot{\bs{q}}_\mu(\bs{x},t)+\bms{\Omega}_\mu^2(\bs{x})\bs{q}_\mu(\bs{x},t) = -\mathbb{Q}^\text{T}(\bs{x}) \cdot \ddot{\bs{u}}(\bs{x},t),
\end{equation}
where $\bms{\Omega}_\mu$ is a diagonal matrix containing the relevant natural frequencies of the inertial subsystem. More details on the derivation of these terms can be found in \citet{Roca2018}.

 
\section{Modelling the viscoelastic behaviour in LRAMs}
\label{sec_metadamping}

The consideration of viscoelastic phenomena affects the model for the stress-strain relation. While in \citet{Roca2018}, materials in the microscale where assumed to behave as linear elastic solids, here, in order to account for rate-dependent effects, a more enriched Kelvin-Voigt model will be introduced (\citet{Krushynska2016,Lewinska2017}), so that the stress-strain relation becomes
\begin{equation}
\label{eq_micro_stress_viscoelastic}
	\bms{\sigma}_\mu(\bs{y},t) = \bs{C}_\mu(\bs{y}) : \nabla_\bs{y}^\text{S}\bs{u}_\mu(\bs{y},t) + \bms{\eta}_\mu (\bs{y}) : \nabla_\bs{y}^\text{S} \dot{\bs{u}}_\mu(\bs{y},t),
\end{equation}  
where $\bs{C}_\mu$ remains as the fourth-order constitutive tensor for an isotropic, linear elastic material and $\bms{\eta}_\mu$ assumes the role of an analogous fourth-order viscous tensor. Since for most polymer-type materials (potential candidates as dissipative coating materials), rate-dependency affects mainly the deviatoric component of the strain velocity, typically the viscosity tensor will be considered as
\begin{equation}
	\bms{\eta}_\mu(\bs{y}) = 2\mu_\mu(\bs{y}) \bs{I}^\text{dev},
\end{equation}
where $\mu_\mu$ is the materials' viscosity distribution and $\bs{I}^\text{dev}$ is the deviatoric fourth-order tensor, defined in index notation as
\begin{equation}
	I_{ijkl}^\text{dev} = \dfrac{1}{2} \left( \delta_{ik}\delta_{jl} + \delta_{il}\delta_{jk} \right) - \dfrac{1}{3} \delta_{ij} \delta_{kl},
\end{equation}
with $\delta_{ij}$ being Kronecker deltas (1 for $i=j$ and 0 otherwise).

Since the hypotheses for homogenization can be compatible with the introduction of this additional effect, the model still holds and the formulation naturally adapts to accommodate this new term. A detailed development of the model's equations with this additional term is provided in \ref{annex_introduction_viscoelastic}, so in this section, only the changes in the results will be discussed, i.e.:
\begin{itemize}
	\item[a)] The first relevant difference is in the expression for the macroscopic stress, which now reads
	\begin{equation}
		\bms{\sigma}(\bs{x},t) \approx \bs{C}^\text{eff}(\bs{x}) : \bms{\varepsilon}(\bs{x},t) + \bms{\eta}^\text{eff} (\bs{x}) : \dot{\bms{\varepsilon}} (\bs{x},t).
	\end{equation}
	Notice the appearance of an effective viscous tensor $\bms{\eta}^\text{eff}$. This is not surprising considering the viscoelastic model assumed for the microscale in equation \eqref{eq_micro_stress_viscoelastic}, which has an analogous form.
	\item[b)] The second difference appears in the projected inertial system of reduced degrees of freedom. While formally the expression for the macroscopic inertia is the same than in equation \eqref{eq_macro_inertia}, an additional damping term appears in former equation \eqref{eq_micro_inertial_system_reduced}, which now reads
	\begin{equation}
	\label{eq_micro_inertial_system_reduced_viscoelastic}
		\ddot{\bs{q}}_\mu(\bs{x},t)+\bms{\Omega}_\mu^\text{D}(\bs{x})\dot{\bs{q}}_\mu(\bs{x},t)+\bms{\Omega}_\mu^2(\bs{x})\bs{q}_\mu(\bs{x},t) = -\mathbb{Q}^\text{T}(\bs{x})\ddot{\bs{u}}(\bs{x},t).
	\end{equation}
	The new matrix $\bms{\Omega}_\mu^\text{D}$ is responsible for \textit{damping} the vibration near the resonance frequencies of the RVE. While this counteracts the effects of local resonance phenomena in the macroscale (especially the higher the frequency becomes), in a relatively low-frequency regime (where LRAMs operate), and for certain levels of viscosity, it can provide the beneficial effect of extending the effective attenuation band. This phenomena will be observed and further described in section \ref{sec_application}.
\end{itemize}

\begin{rmk}
	It is important to notice that, while the system of equations \eqref{eq_micro_inertial_system_reduced} is fully uncoupled, which allows us to perform an effective reduction of the number of degrees of freedom that need to be considered in the analysis, the same cannot be guaranteed for the system \eqref{eq_micro_inertial_system_reduced_viscoelastic}, due to the presence of the matrix $\bms{\Omega}_\mu^\text{D}$, which is non-diagonal, in general, and can make the homogenization model more computationally expensive than in the case where no viscoelastic effects are considered. This increase in computational cost is related to the degree of copuling existing in the matrix $\bms{\Omega}_\mu^\text{D}$, which ultimately depends on the RVE topology, material properties and modelling of viscoelastic effects in the microscale. For instance, \old{for a proportional Rayleigh damping model (in which the damping matrix is proportional to the stiffness matrix), it is easy to see that $\bms{\Omega}_\mu^\text{D}$ would become diagonal (since it would be proportional to $\bms{\Omega}_\mu^2$), hence, in this case,} the system \eqref{eq_micro_inertial_system_reduced_viscoelastic} would remain fully uncoupled \new{only in cases where the damping matrix is already diagonal or either it is proportional to the stiffness and/or the mass matrices (typically known as proportional Rayleigh damping model), which cause the matrix $\bms{\Omega}_\mu^\text{D}$ to be diagonal. In the specific cases accounting for viscous effects considered in this work, strain rate dependence is only considered in one of the material components, i.e. the coating, thus, in general, the matrix $\bms{\Omega}_\mu^\text{D}$ is non-diagonal. However, the degree of coupling with non-relevant modes will be small enough to be negligible in practice.}.
\end{rmk}

\section{Topological design of LRAMs}
\label{sec_topology}
\begin{figure*}
	\centering
	\includegraphics{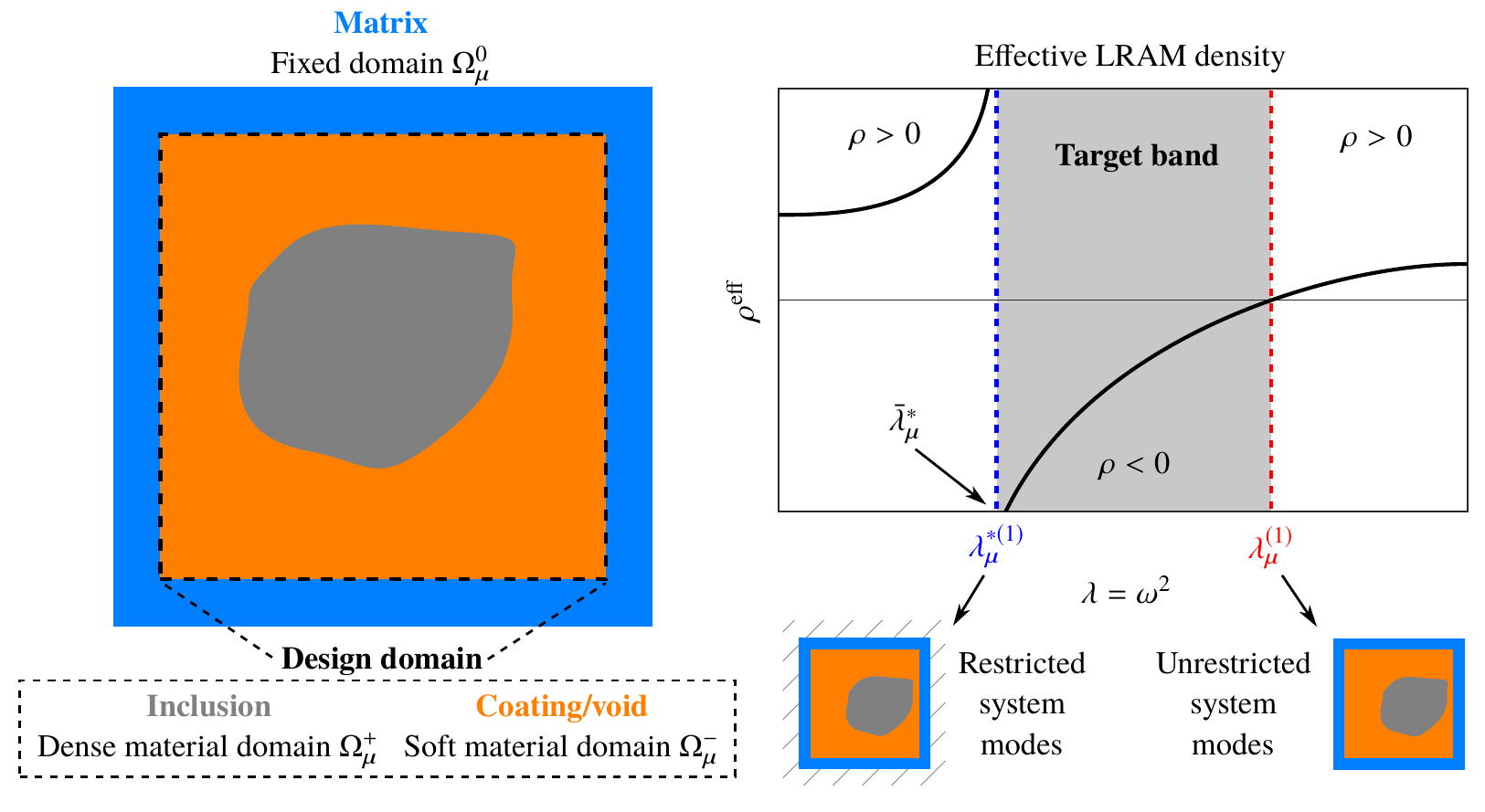}
	\caption{RVE configuration for the LRAM topology optimization (left) and typical effective LRAM density diagram depicting the frequency bandgap corresponding to negative densities (right). In order to ensure LRAM-like behaviour, a fixed matrix of material is considered, $\Omega_\mu^0$, with constant material properties and $\chi(\bs{y}) = 1, \forall \bs{y} \in \Omega_\mu^0$. Only in the inside region the characteristic function $\chi$ is allowed to change giving rise to \textit{dense material} volumes corresponding to inclusions, $\Omega_\mu^+$, and \textit{soft material} volumes corresponding to void/coating material, $\Omega_\mu^-$. The resulting resonance frequencies $\omega_\mu^{*(1)} = \sqrt{\lambda_\mu^{*(1)}}$ and $\omega_\mu^{(1)} = \sqrt{\lambda_{(1)}}$ associated to the \textit{restricted} and \textit{unrestricted} system modes, respectively, determine the lower and upper bounds of the target band, whose lower bound is matched to $\bar{\lambda}_\mu^{*}$.}
	\label{fig1}
\end{figure*}

Aiming at obtaining LRAMs with topologies designed to achieve better attenuation properties, i.e. increasing the levels and range of transmission loss (in the specific case of acoustic insulation panels), a level-set based topology optimization strategy is proposed here with two main objectives:
\begin{itemize}
	\item[a)] Fit the relevant resonance frequencies of the resulting topology into a targeted band. This is done by matching the lower-bound of the target band with a relevant resonance frequency of the RVE (see Figure \ref{fig1}).
	\item[b)] Maximize the bandwidth of the target band in terms of the topology of the RVE materials. 
\end{itemize}
In such methodologies, a suitable cost function is minimized, in a variational way, with respect to a characteristic function 
\begin{equation}
	\chi(\bs{y}) : \Omega_\mu \rightarrow \{0, 1\},
\end{equation}
that defines the material distribution in the design domain $\Omega_\mu$, taking values of 1 for \textit{dense material} regions (inclusions): $\bs{y}\in\Omega_\mu^+$ and 0 for \textit{soft material} regions (coating/void): $\bs{y}\in\Omega_\mu^-$ ($\Omega_\mu^+ \cup \Omega_\mu^- = \Omega_\mu$). These regions are typically determined by a \textit{smooth} level-set function $\phi(\bs{y})$ such that
\begin{equation}
\label{eq_chi}
	\chi(\bs{y}) := \mathcal{H}(\phi(\bs{y})) \equiv \left\lbrace
	\begin{matrix}
	0 \quad \forall \bs{y} \ \text{ such that } \ \phi(\bs{y}) < 0\\
	1 \quad \forall \bs{y} \ \text{ such that } \ \phi(\bs{y}) \ge 0
	\end{matrix}
	\right. ,
\end{equation}
so that the function $\phi(\bs{y})$ becomes the unknown of the problem in a variational context.

Prior to presenting the optimization problem itself, let us first review the RVE equations that will be required to solve this problem. First, in order to obtain the RVE's relevant resonance frequencies, the modal problem of the \textit{restricted} system must be solved
\begin{equation}
\label{eq_modal_restricted}
	( \mathbb{K}_\mu^\ast - \lambda_\mu^{*(k)} \mathbb{M}_\mu^\ast ) \hat{\bms{\phi}}_\mu^{*(k)} = \bs{0}, \quad \hat{\bms{\phi}}_\mu^{*(k)\text{T}} \mathbb{M}_\mu^\ast \hat{\bms{\phi}}_\mu^{*(k)} = 1,
\end{equation} 
where $\mathbb{K}_\mu^\ast$ and $\mathbb{M}_\mu^\ast$ are the resulting stiffness and mass matrices once the kinematic restrictions on the microfluctuation field have been applied. In this specific case, since the local resonance phenomenon occurs at frequencies corresponding to \textit{internal} vibration modes, a good approximation to meet our goals consists of prescribing all RVE boundaries. The terms $\lambda_\mu^{*(k)}$ and $\hat{\bms{\phi}}_\mu^{*(k)}$ correspond to the squared natural frequencies and mass-normalized vibration modes of the system. According to equation \eqref{eq_macro_inertia}, only those modes $\hat{\bms{\phi}}_\mu^{*(k)}$ such that their corresponding columns in the generalized coupling matrix $\mathbb{Q}$ are larger than a certain tolerance $\delta_\text{tol}$, should be considered as relevant. Thus, the set of \textit{relevant} resonance frequencies is given by
\begin{equation}
	\omega_\mu^{*(k)} = \sqrt{\lambda_\mu^{*(k)}} \quad \text{such that} \quad \|\mathbb{Q}^{(k)}\| \equiv \|\avg{\rho_\mu \bms{\phi}_\mu^{*(k)}}_{\Omega_\mu} \| > \delta_\text{tol}.
\end{equation}
Furthermore, since transmission loss peaks in LRAM panels are closely related to frequency bandgaps, whose lower and upper limits can be identified, respectively, with the relevant resonance frequencies of the \textit{restricted} and \textit{unrestricted} RVE system, $\omega_\mu^{*(k)}$ and $\omega_\mu^{(k)}$, as reported in \citet{Roca2018} (see Figure \ref{fig1} for a schematic representation), one can maximize their bandwidth, for instance, by minimizing the ratio $\lambda_\mu^{*(k)}/\lambda_\mu^{(k)}$, where $\lambda_\mu^{(k)}$ comes from the modal problem considering the mass and stiffness matrices of the RVE system prior to applying the kinematic restrictions, i.e.
\begin{equation}
\label{eq_modal_unrestricted}
( \mathbb{K}_\mu - \lambda_\mu^{(k)} \mathbb{M}_\mu ) \hat{\bms{\phi}}_\mu^{(k)} = \bs{0}, \quad \hat{\bms{\phi}}_\mu^{(k)\text{T}} \mathbb{M}_\mu \hat{\bms{\phi}}_\mu^{(k)} = 1,
\end{equation} 
where in this case, the only relevant $\lambda_\mu^{(k)}$ that correspond to upper bounds of the frequency bandgaps can be identified by
\begin{equation}
	\omega_\mu^{(k)} = \sqrt{\lambda_\mu^{(k)}} \quad \text{such that} \quad \lambda_\mu^{(k)} > 0 \ \text{ and } \ \|\avg{\bms{\phi}_\mu^{(k)}}_{\Omega_\mu}\| > \delta_\text{tol}.
\end{equation}
Note that, in this case, the additional condition $\lambda_\mu^{(k)} > 0$ needs to be applied in order to avoid rigid body translation modes, which are relevant according to condition $\|\avg{\bms{\phi}_\mu^{(k)}}_{\Omega_\mu}\| > \delta_\text{tol}$, but do not define the upper bounds of any bandgap.

\begin{figure*}
	\centering
	\includegraphics{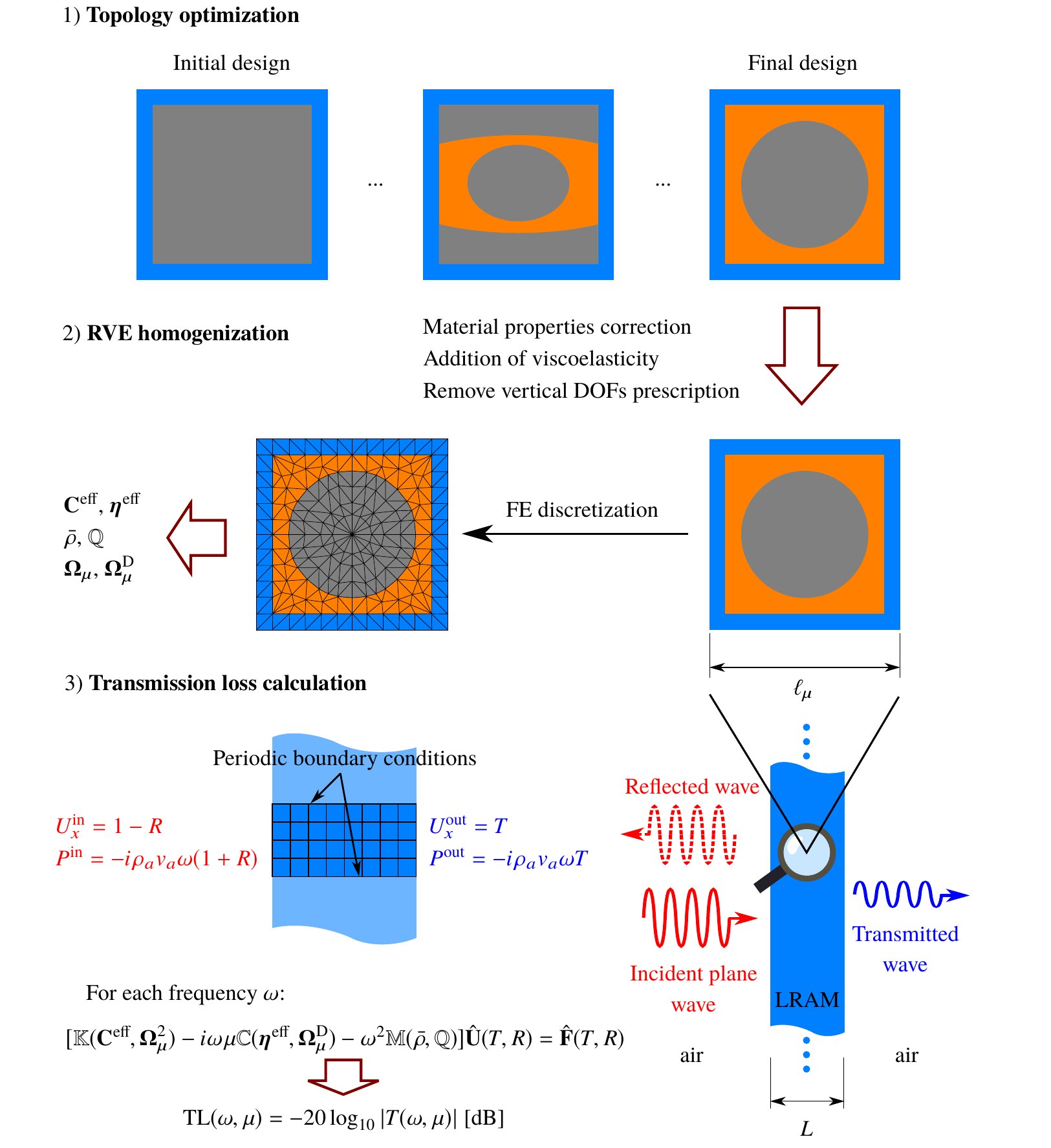}
	\caption{Global procedure to evaluate the transmission loss of a LRAM panel with a topology optimized design. The material distribution is obtained from the topology optimization algorithm. This result is used to build a RVE with the actual material properties from which the effective properties are computed by employing the multiscale homogenization framework. Finally, a macroscale analysis is performed over a slice of the panel, imposing displacement and traction compatibility conditions with the incoming and outgoing waves and periodic boundary conditions at the material boundaries (in order to simulate the infinite extension of the panel in the vertical direction). This analysis is performed in the frequency domain for several test frequencies allowing us to obtain the transmission and reflection coefficients of the panel, $T$ and $R$, respectively.}
	\label{fig2}
\end{figure*} 

In this regard, the objective function proposed to minimize is given by
\begin{equation}
\label{eq_objective_fcn}
\Pi(\chi(\phi)) = \alpha f^2 + (1-\alpha) g^2,
\end{equation}
with
\begin{align}
\label{eq_objective_fcn_1}
&f = \dfrac{\ln \lambda_\mu^{*(1)}(\chi(\phi)) - \ln \bar{\lambda}_\mu^{*}}{\ln \lambda_\mu^{*(1)}(\chi(\phi)) + \ln \bar{\lambda}_\mu^{*}}, \\
\label{eq_objective_fcn_2}
&g = \dfrac{\ln \lambda_\mu^{*(1)}(\chi(\phi))}{\ln \lambda_\mu^{(1)}(\chi(\phi))},
\end{align}
subject to the state-equations \eqref{eq_modal_restricted} and \eqref{eq_modal_unrestricted}. In equations \eqref{eq_objective_fcn} to \eqref{eq_objective_fcn_2}, $\alpha$ is a weighting parameter to establish the relative importance of each term in the global objective function, $\bar{\lambda}_\mu^{*}$ is the imposed targeted squared frequency to fit with the first relevant squared resonance frequency for the \textit{restricted} RVE system $\lambda_\mu^{*(1)}$, while $\lambda_\mu^{(1)}$ refers to the first relevant squared resonance frequency for the \textit{unrestricted} RVE system. Note that, as long as $0 \leq \alpha \leq 1$, the objective function will be bounded $\Pi \in [0,1]$, so that it reaches its minimum value when all the desired objectives are fulfilled. For the sake of simplicity, only the first frequency band is targeted to fit in this framework, but equation \eqref{eq_objective_fcn} can be easily extended to target multiple frequency bands simply by adding the corresponding terms in the cost function.

The optimization problem then reads
\begin{align}
	&\text{FIND:}& &\chi(\bs{y})=\mathcal{H}(\phi(\bs{y})): \Omega_\mu \rightarrow \{0, 1\};& & & & & & & & \nonumber\\
	&\text{FULFILLING:}& &\chi = \text{arg } \min_{\chi} \Pi(\lambda_\mu^{*(1)}(\chi),\lambda_\mu^{(1)}(\chi)),& & & & & & & & \nonumber\\
	& & &\text{s.t.} & &\\
	& & &( \mathbb{K}_\mu^\ast - \lambda_\mu^{*(1)} \mathbb{M}_\mu^\ast ) \hat{\bms{\phi}}_\mu^{*(1)} = \bs{0},& & & & & & & & \nonumber\\
	& & &( \mathbb{K}_\mu - \lambda_\mu^{(1)} \mathbb{M}_\mu ) \hat{\bms{\phi}}_\mu^{(1)} = \bs{0}.& & & & & & & &\nonumber
\end{align}
A time-marching technique is used to update the problem with a pseudo-time variable, from an initial state (typically all the design area full with dense material) towards a problem's solution. In this case, a Hamilton-Jacobi approach has been considered in which the problem's evolution has been defined in a rate form as
\begin{equation}
\label{eq_HJac_rate}
	\dot{\phi}(\bs{y},t) \equiv \dfrac{\partial \phi(\bs{y},t)}{\partial t} = - C_1 \dfrac{\delta \Pi_t}{\delta \chi}(\bs{y}),
\end{equation}
allowing us to obtain the updated value of function $\phi$ from the previous iteration step through a straightforward time discretization of equation \eqref{eq_HJac_rate}
\begin{equation}
\label{eq_HJac_rate_discrete}
	\phi^{n+1}(\bs{y}) = \phi^n(\bs{y}) - \Delta t C_1 \dfrac{\delta \Pi_t}{\delta \chi}(\bs{y}), \quad \forall\bs{y}\in\Omega_\mu
\end{equation}
where $\Delta t$ is a pseudo-time step, $C_1>0$ is a parameter and $\delta\Pi_t/\delta\chi$ is the topological sensitivity, evaluated at point $\bs{y}$, of the cost function \eqref{eq_objective_fcn}, here named as the Variational Topological Derivative (VTD) of the functional $\Pi$. In \ref{app_sensitivity_cost_fcn} it is proven that the iterative scheme in equation \eqref{eq_HJac_rate_discrete} yields an iterative descend of the cost function $\Pi$, i.e. $\dot{\Pi}_t(\chi) \leq 0 \ \forall t$, a crucial aspect for the convergence of the Hamilton-Jacobi algorithm.

To ensure that local resonance phenomena arise in the computed designs throughout the optimization process, the frequency fitting will always be required, and typically aimed at the smallest relevant resonance frequency achievable (i.e. $\alpha>0$) which, for a given set of material properties, will be constrained by the dimensions of the design domain (in this case the RVE). Furthermore, the RVE will consist of a fixed matrix material frame (non-design domain) so that the actual design domain is the inclusion/coating distribution on the inside (see Figure~\ref{fig1}).

Some additional hypothesis have been made in order to avoid spurious modes resulting from the modal analysis, which helps the optimization algorithm to become more stable and converge to the desired solutions. These hypothesis are listed and explained below:
\begin{itemize}
	\item[a)] The matrix fixed frame is considered infinitely stiff so that no deformation modes, which are non-relevant in this context, appear in the modal analysis. This is done to prevent modes in which the matrix interacts with the other materials, especially in early steps of the algorithm, when the whole domain is filled with material.
	\item[b)] The void/coating material is considered massless. Since the modes causing the local resonance phenomena to arise are those in which the inclusions vibrate inside the coating phase, the relevant properties that are to be considered will be the density of the inclusion phase and the stiffness of the coating material. Forcing the density of the coating material to zero (or a very small tolerance value), avoids the appearance of spurious modes in the modal analysis which greatly helps the identification of the relevant resonance modes.
	\item[c)] Since the focus in this context is in horizontally-oriented modes (the panel will be subjected to plane waves propagating on the horizontal direction), all vertical degrees of freedom are prescribed in both the restricted and unrestricted systems. By restricting the analysis to a single dimension, both the identification of the relevant modes and the pairing of each bandgap limiting frequencies become much easier.   
\end{itemize}

\begin{table}
\centering
\caption{Material properties \citep{Calius2009}.}
\label{tab1}
\begin{tabular}{lccc}
	\hline
	\multirow{2}{*}{Material} &   Density    &   Bulk mod.   &  Shear mod.   \\
	                          & $\rho_\mu$ (kg/m${}^3$) &    $K_\mu$ (MPa)       &   $G_\mu$   (MPa)       \\ \hline
	Epoxy                     &     1180     & $5.49\times10^3$ & $1.59\times10^3$ \\
	Steel                     &     7780     & $1.72\times10^5$ & $7.96\times10^4$ \\
	Silicone rubber           &     1300     &       0.63       &       0.04       \\ \hline
\end{tabular} 
\end{table}

\begin{figure}
	\centering
	\includegraphics{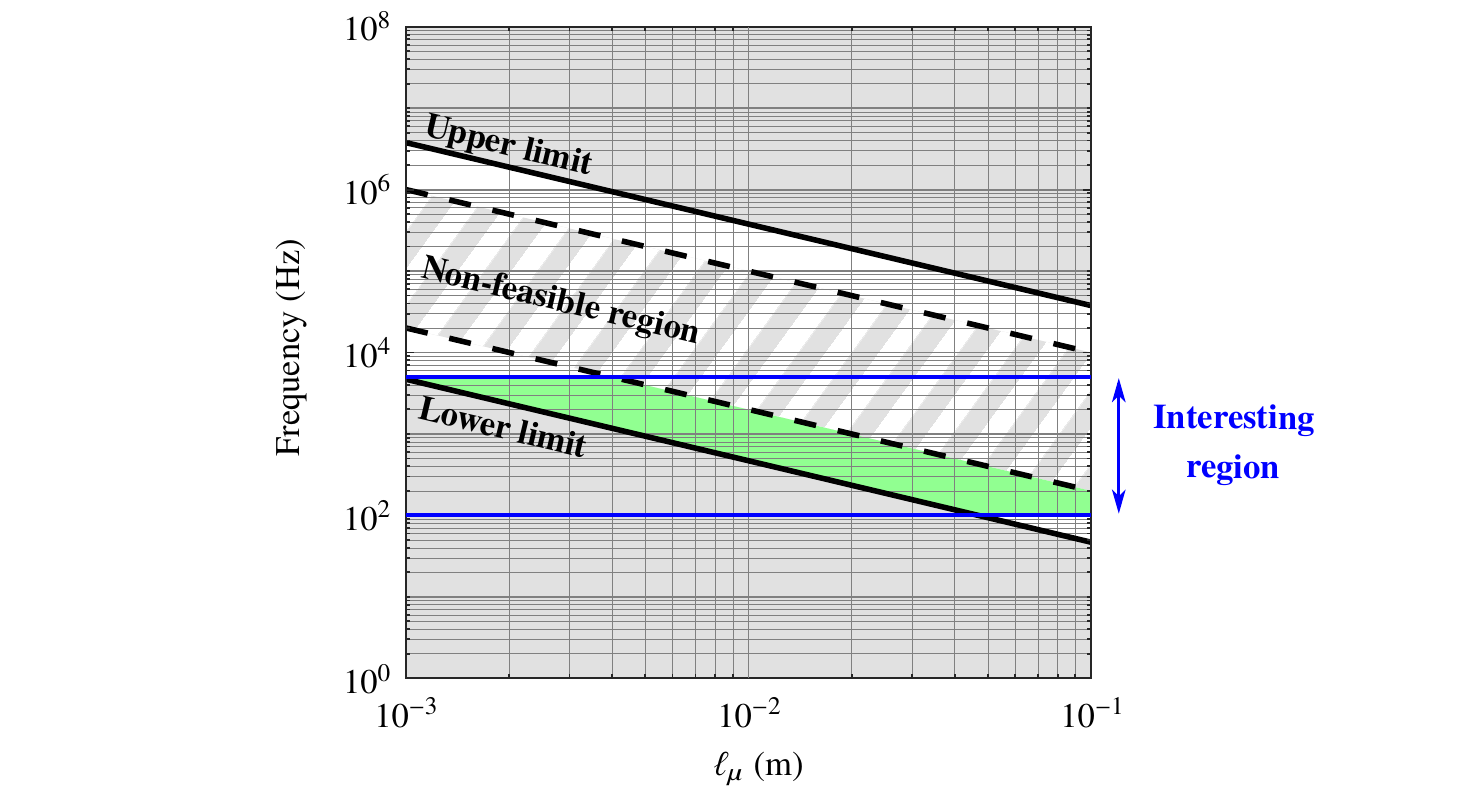}
	\caption{Representation of the valid regions for achievable first target resonating frequencies in terms of the RVE size. Upper and lower limits and the so-called \old{numerical gap}\new{non-feasible region} have been obtained with the material and numerical properties for this example, which are listed in Table \ref{tab1}. The \textcolor[rgb]{0.25,1,0.25}{green} shaded area corresponds to the region of achievable first target resonating frequencies within the range of interest.}
	\label{fig3}
\end{figure} 

\section{Application to the design of an acoustic insulation panel}
\label{sec_application}
Let us consider an infinitely large flat panel with a given thickness $L$ built with stacked LRAM unit cells (of size $\ell_\mu$) consisting of 3 material phases: an epoxy matrix frame at the boundaries along with a certain distribution of steel inclusions embedded in a silicone rubber coating (see Figure \ref{fig1} for a graphical depiction of a typical RVE configuration). The material properties used in the examples that follow are listed in Table \ref{tab1}. The coating material will be considered viscoelastic, with the viscosity $\mu$ left out as a parameter in order to enable the possibility of evaluating the LRAM behaviour for various degrees of dissipation. Since the aim of this analysis is to assess the attenuation of acoustic waves through a slab of a designed LRAM panel, the transmission loss in the specific frequency range of interest (in this case below 3000 Hz) will be computed. To do so, a 3-step analysis is performed. First, the topology optimization procedure explained in section \ref{sec_topology} is used to obtain a material distribution for the LRAM design that meets the desired properties. Then, a RVE is built upon the results obtained from the previous step with the actual material properties, so the homogenization procedure detailed in sections \ref{sec_LRAM_homog} and \ref{sec_metadamping} can be applied to compute the effective material properties associated to that LRAM design. Finally, an analysis on the macroscale is performed by computing the transmission loss in the desired frequency range for a flat panel composed of the homogenized LRAM  subjected to acoustic plane waves. Details on each of these procedures is given in the following sections, while a summary of the global scheme is depicted in Figure \ref{fig2}. \new{A plane-strain 2D approach is considered in all the examples instead of a 3D setting simply to avoid the unnecessary complexity associated with them which, at least regarding the effects and conclusions that are expected to point out in this work, do not give any relevant additional insights. Therefore, for the sake of clarity in terms of interpretation of the results, the examples shown here are all 2D, even though both the formulation and the conclusions that can be extracted can be extended to 3D.}

\subsection{Topology optimization of the LRAM}
\label{sec_topology_optim_LRAM}

\begin{figure*}
	\centering
	\includegraphics{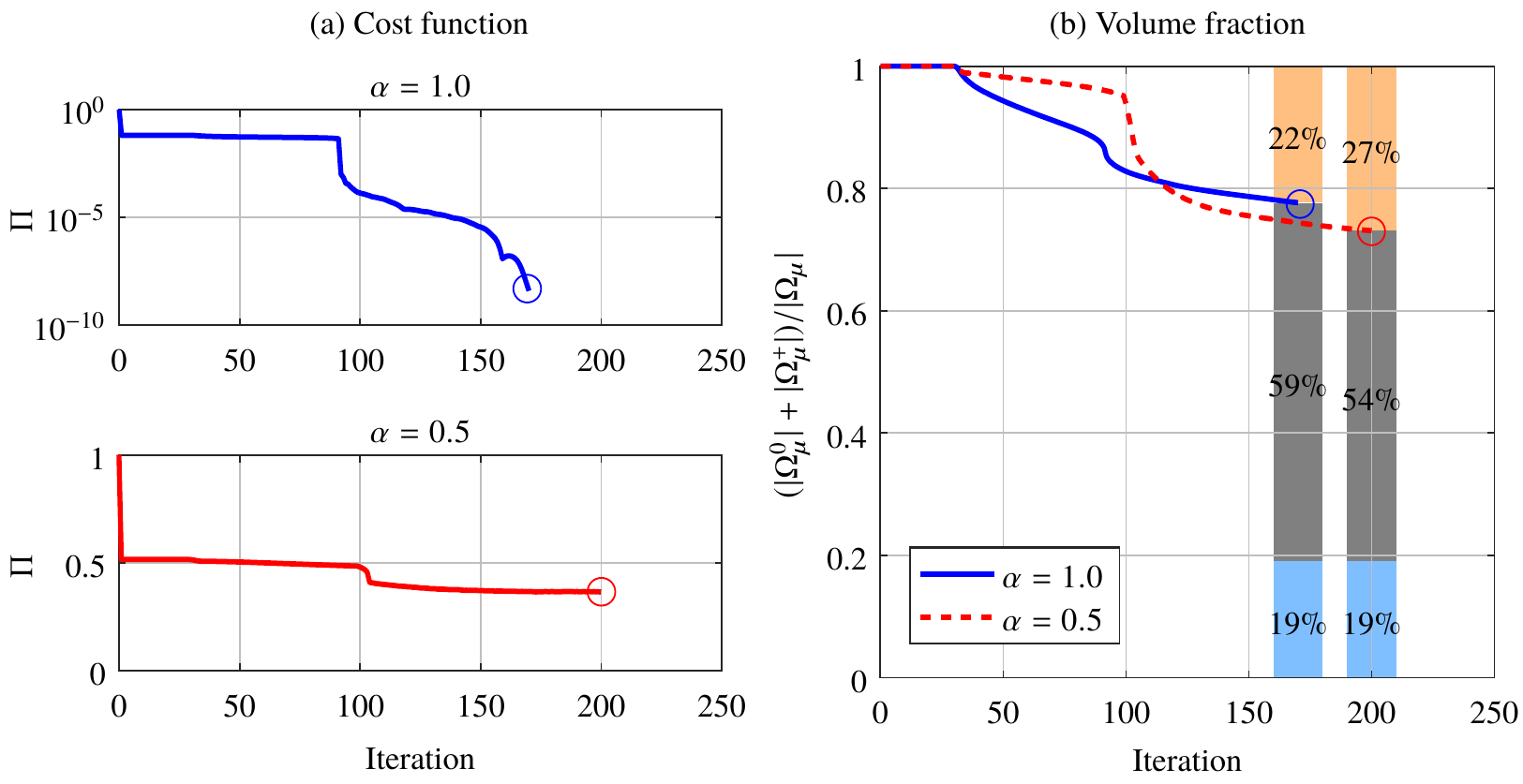}
	\caption{(a) Evolution of the objective function upon each iteration for the frequency fitting case (top) and the frequency fitting along with bandgap maximization (bottom). For $\alpha = 0.5$, a trade-off must be met between the frequency fitting part and the bandgap maximization component, which makes it more difficult for the algorithm to converge. (b) Evolution of the volume fraction upon each iteration. Note that the colorbar represents the volume fraction distribution of each material in the domain. The fixed matrix (\textcolor[rgb]{0.5,0.75,1}{blue}) represents 19\% of the volume in both cases, and the remaining 81\% is distributed between inclusion (\textcolor[rgb]{0.5,0.5,0.5}{grey}) and coating phases (\textcolor[rgb]{1,0.75,0.5}{orange}).}
	\label{fig4}
\end{figure*} 


\begin{figure*}
	\centering
	\includegraphics{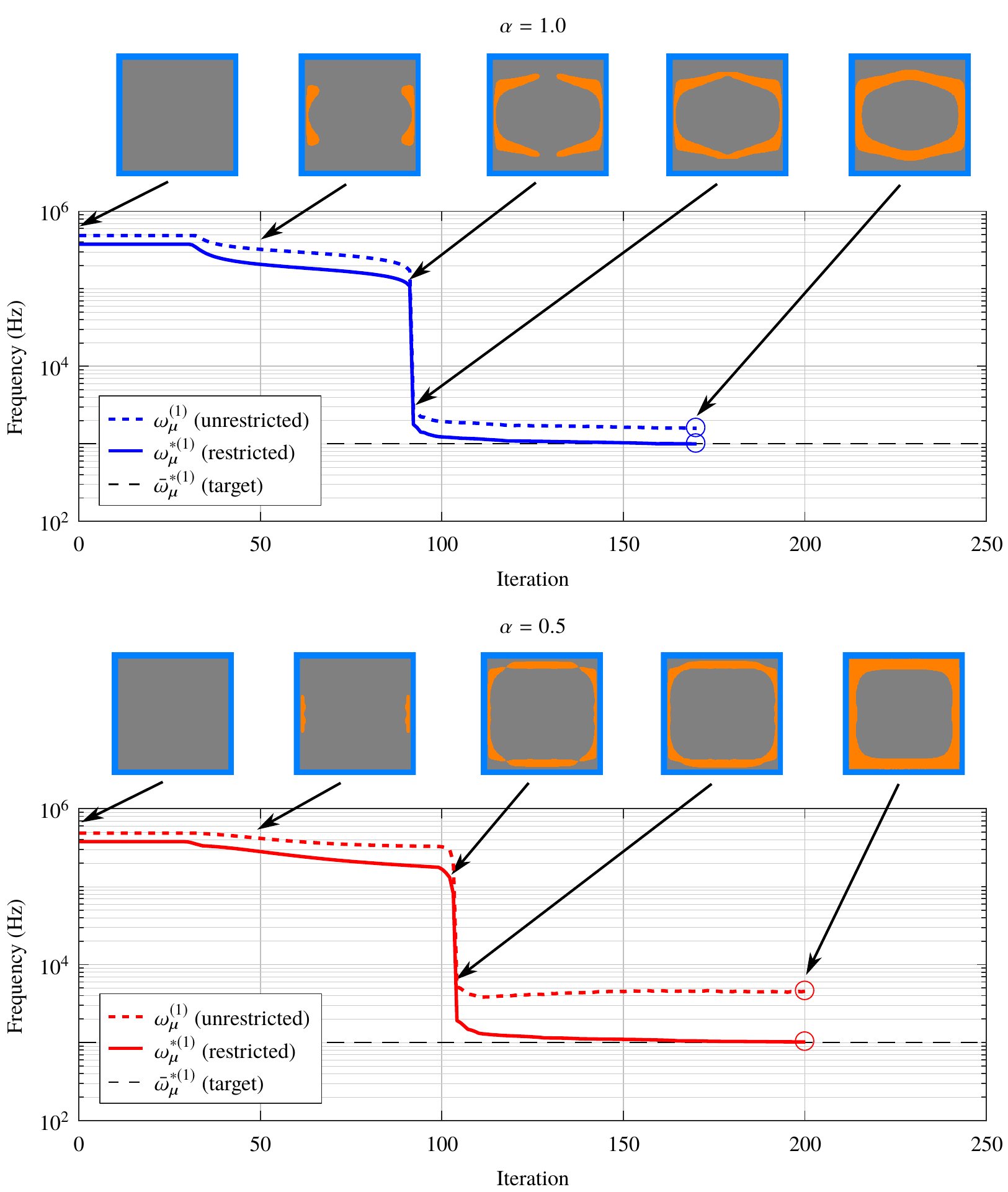}
	\caption{Evolution of the first relevant resonance frequencies for the restricted and unrestricted systems upon each iteration for the frequency fitting case (top) and frequency fitting along with bandgap maximization case (bottom). The topology  determined by the level-set function $\phi$ at several iteration steps is also shown. Note the sudden jump in frequencies coincides with the iteration when the inclusion is disengaged. This gap gives an idea of the region of unattainable frequencies due to numerical issues.}
	\label{fig5}
\end{figure*}
 	
\begin{figure*}
	\centering
	\includegraphics{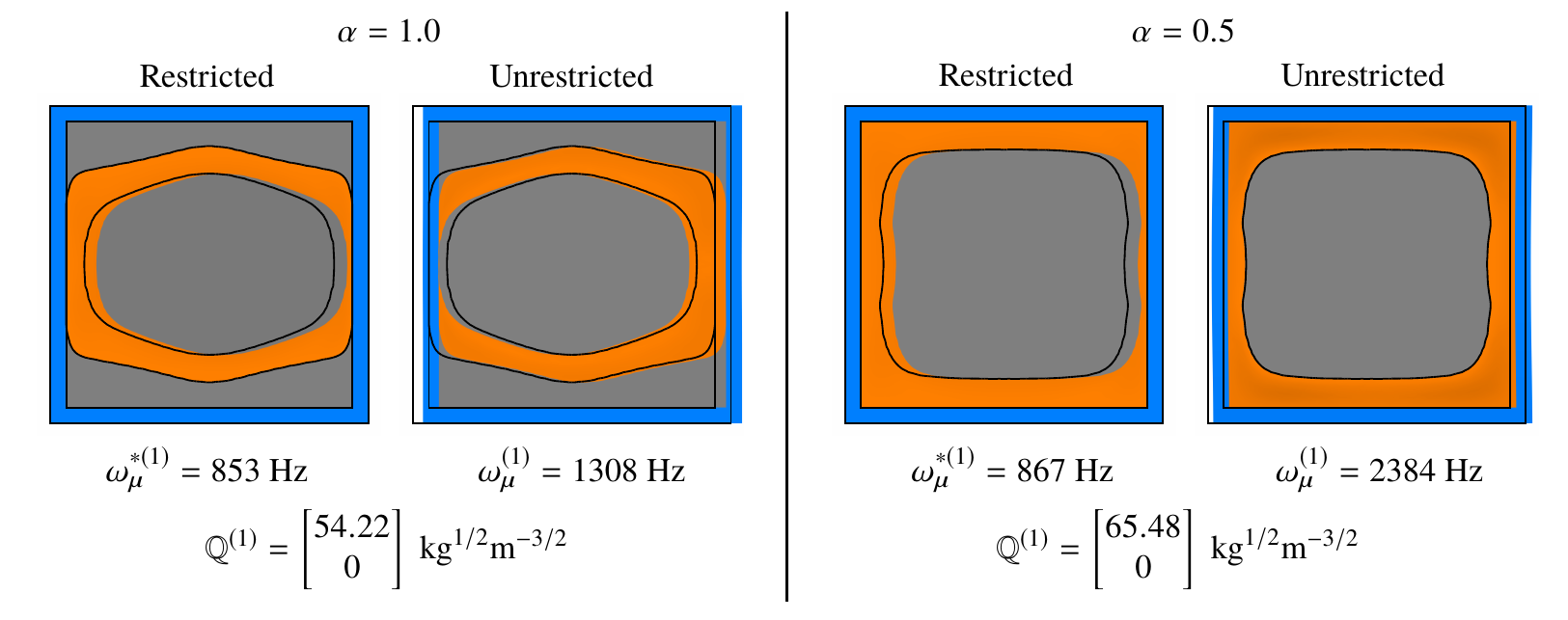}
	\caption{First relevant resonance modes and frequencies for the restricted and unrestricted systems for the topologies resulting from the frequency fitting optimization (left) and the frequency fitting coupled with the bandgap maximization (right). The solid black lines represent the material interfaces of the non-deformed state.}
	\label{fig6}
\end{figure*}

A 2D structured mesh of $100\times100$ quadrilateral elements with 4 Gaussian integration points has been used for the computations in this stage. The elastic properties of the matrix phase have been scaled by a factor $10^{10}$ to guarantee its behaviour as a rigid component, while the density of the coating has been scaled by $10^{-10}$ to avoid spurious, non-relevant modes resulting from the modal analysis. A value of $\Delta t = 10^{-3}$ has been considered as a pseudo-timestep for the time-marching algorithm, along with an initial $\phi^0 > 0$ that makes all the design domain to be full of inclusion dense material (see equation \eqref{eq_chi}). The size of the RVE has been chosen to be 1 $\times$ 1 cm to ensure that the resulting resonating frequencies lie on the desired range. This is important since for a given set of material properties and a domain size, there are limitations in the achievable resonant frequencies. An obvious first upper limit can be found in the first resonant frequency obtained by the initial full-material configuration, which depends on the properties of the inclusion phase as well as the RVE size and is typically well above (by several orders of magnitude) the desired frequency range. On the other hand, a theoretical lower limit can also be estimated as
\begin{equation}
	\omega_\mu^{*(1)} > \dfrac{1}{\ell_\mu} \sqrt{\dfrac{\min\limits_i \left\{ K_\mu^{(i)}+\dfrac{4}{3}G_\mu^{(i)} \right\} }{\max\limits_i \rho_\mu^{(i)}}}
\end{equation} 
where the index $i$ here refers to each material. Furthermore, given the high contrast between the properties of the inclusion and coating components, there is a range of frequencies between these lower and upper limits for which the algorithm finds it more difficult to converge to a solution. This is because solutions in this range typically contain \new{physically unstable solutions characterised by large changes in the resonance frequencies for small perturbations in the topology (in this case, caused by the appearance of unrealistically thin strings of material). Therefore, designs inside this zone should be avoided to preclude such unstable behaviour. Note also that these solutions } 
\old{too small features which require both a very fine mesh capable of capturing them and also a very small timestep for not ignoring them during the algorithm. In practice, solutions in this range are not interesting from a manufacturing point of view, and} can be easily avoided for a given target frequency by reducing the size of the design domain (see Figure \ref{fig3} as an example).

For this example, a target frequency of $\bar{\omega}_\mu^{*(1)}=1000$ Hz is selected and two different weighting parameters have been tested: one that causes only the frequency fitting to be considered in the objective function ($\alpha=1$), and another one where the weight is distributed equally among the fitting and the bandgap maximization parts ($\alpha=0.5$). Figure \ref{fig4} shows the evolution of the objective function and the volume fraction of material, respectively. The first relevant resonating frequencies for the restricted and unrestricted problems can be seen in Figure \ref{fig5}. 

These results show how the algorithm removes inclusion material in the topology, reducing the value returned by the objective function, as seen Figure \ref{fig4} (a), and thus approaching the desired targets with each iteration step. It is worth noting the sudden jump produced around the 100th iteration in both cases. It is caused by the disengagement of the inclusion material from itself, which makes the coating phase to fully envelop it. Since the stiffness of the coating material is several orders of magnitude lower, the resulting configuration makes it easier for the enclosed inclusion to vibrate, which translates into a much lower resonance frequency. 

Interestingly, while the volume fraction of material ends up being very similar in both cases, as seen in Figure \ref{fig4} (b), the resulting topologies are quite different. Note also that the resonance frequency for the restricted problem follows a similar evolution in both cases (eventually meeting the target frequency of 1000 Hz), but the differences between their associated topologies make their respective unrestricted system's resonance frequency to evolve differently (see Figure \ref{fig5}). In particular, when the bandgap maximization is also part of the objective function the algorithm tends to remove material from the matrix internal borders, concentrating the maximum amount of mass onto the central inclusion. This translates into a significant increase of the bandgap size, for a similar volume fraction of material distribution, by almost 6 times (around 600 Hz in the first case and up to 3500 Hz in the second).

\begin{rmk}
	\label{rmk_freq}
	It must be considered that the frequencies obtained from the optimization process will not be equal to those computed when actual material properties (without scaling factors) are considered, as it will be seen in the results of the next section. However, this is not a major issue for our purposes, since with the actual material properties, the system tends to be less restricted and so the resulting frequencies are smaller, which is often desirable. In any case, the size of the RVE can be adjusted accordingly to match the target frequencies, if desired.
\end{rmk}

\subsection{Homogenization of the optimized LRAM}

\begin{figure*}
	\centering
	\includegraphics{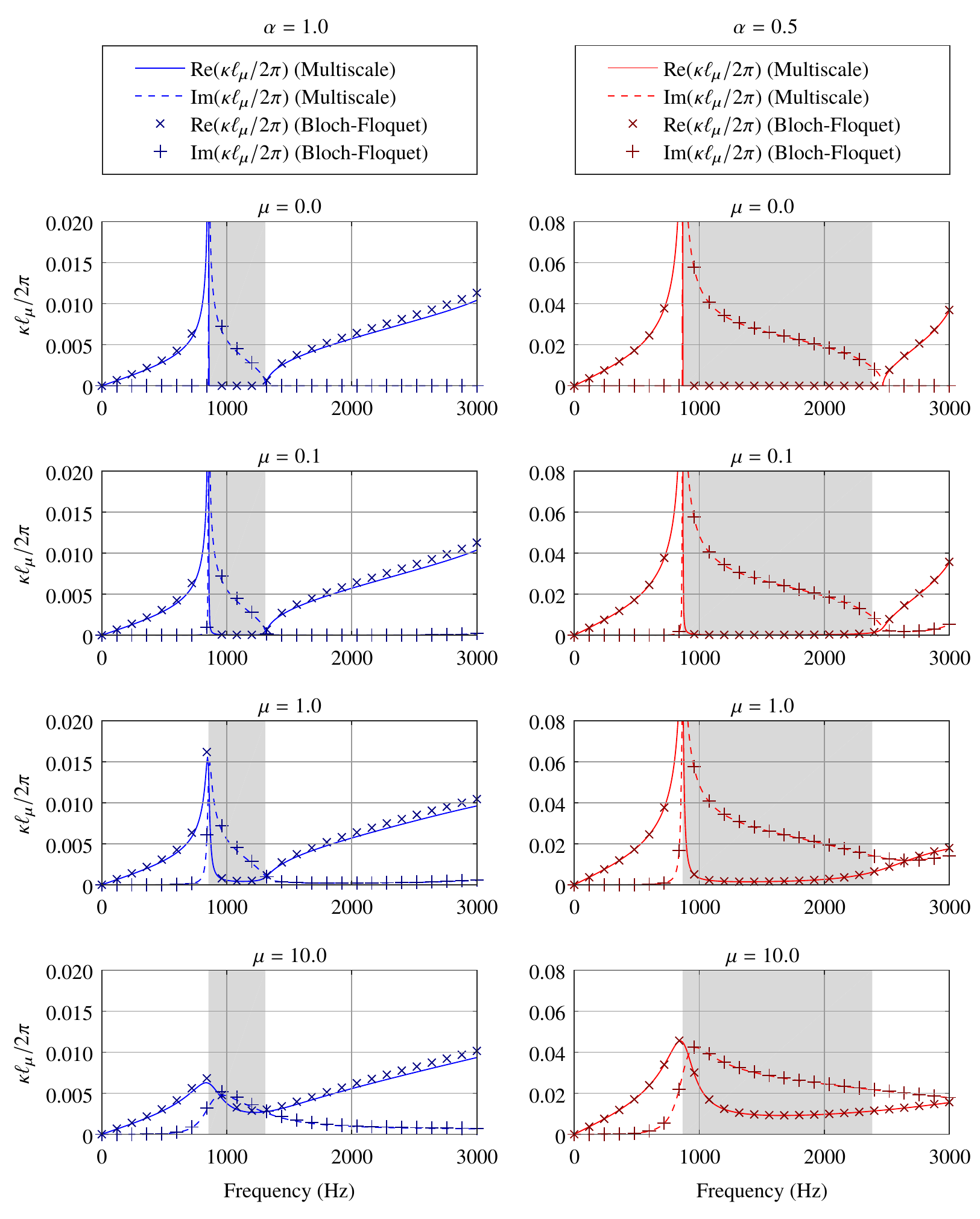}
	\caption{Dispersion diagram for the topologies resulting from the frequency fitting optimization (left) and the frequency fitting coupled with the bandgap maximization (right). The shaded areas correspond to the bandgap regions corresponding to purely imaginary wavenumbers in the case without viscosity ($\mu=0$). The units for the viscosity parameter $\mu$ are Pa$\cdot$s. Note that the real part of the normalized wavenumber corresponds to the solid lines and 'x' markers, while its imaginary component is represented through dashed lines and '+' markers in the same axes. The results obtained with the multiscale homogenization procedure are compared with those obtained applying Bloch-Floquet boundary conditions on the same RVEs, for validation purposes.}
	\label{fig7}
\end{figure*}

With the material distributions obtained in the previous step, RVEs are built, maintaining their size but with the actual material properties. The homogenization framework detailed in sections \ref{sec_LRAM_homog} and \ref{sec_metadamping} is applied to compute the effective material parameters associated to each LRAM design. These include the effective constitutive tensor, $\bs{C}^\text{eff}$, the average density $\bar{\rho}$, the coupling matrices related to micro-inertial phenomena, $\mathbb{Q}$, $\bms{\Omega}_\mu$, and the additional terms $\bms{\eta}^\text{eff}$, $\bms{\Omega}_\mu^\text{D}$, accounting for viscous effects. For these computations, 2D meshes of triangle elements with an average relative size of 0.02 and 3 Gaussian integration points have been used, which provide around 7000 degrees of freedom. Figure \ref{fig6} shows the resulting first resonance modes and frequencies which, as previously anticipated in Remark \ref{rmk_freq}, are smaller than the computed as part of the optimization process in the previous section. Note, however, that the predicted bandgap in the case with $\alpha=0.5$ is much larger than in the other case (around 1500 to 450 Hz, which is between 3 and 4 times bigger). This can also be seen in the dispersion diagrams of Figure \ref{fig7}, which are calculated assuming a plane wave travelling in the $x$-direction in an infinite extension of the homogenized material domain (see \cite{Roca2018} for details on this computation). For the case without viscosity ($\mu=0$), a non-null imaginary component of the wavenumber indicates the presence of a frequency bandgap. In Figure \ref{fig7}, the results of applying Bloch-Floquet boundary conditions on the same RVEs \citep{Hussein2014} are also shown to validate the proposed model. Note that there is good agreement between both computations, even for the cases where viscosity is present.

\subsection{Transmission loss computation}
Finally a macroscale analysis is performed over a homogenized panel from which the transmission coefficient is obtained. For the sake of simplicity, the panel is assumed to be surrounded by air (with density $\rho_a = 1.2 \ \text{kg/m}^3$ and sound propagation speed $v_a = 344 \ \text{m/s}$) where a plane wave travels perpendicular to the panel's surface. In order to find the transmission coefficient in this case, compatibility conditions for the normal component of the displacement and traction forces are imposed at the panel's interfaces with the air media. In this context, one can assume the air at the left hand side to propagate as a wave at a certain frequency $\omega$:
\begin{align}
\label{eq_comp1}
	&u_x^{(l)}(\bs{x},t) = (e^{i\kappa_a x} - Re^{-i\kappa_a x})e^{-i\omega t}, \\
\label{eq_comp12}
	&p^{(l)}(\bs{x},t) = -i\rho_a v_a \omega (e^{i\kappa_a x} + R e^{-i\kappa_a x})e^{-i\omega t} 
\end{align}
where $u_x^{(l)}$ and $p^{(l)}$ are the horizontal displacement and pressure fields, respectively, $\kappa_a = \omega/v_a$ is the wavenumber and $R$ is the reflection coefficient, i.e. the fraction of the incident wave's amplitude that is reflected at the panel's interface. At the right hand side, only part of the wave is transmitted, so
\begin{align}
\label{eq_comp21}
&u_x^{(r)}(\bs{x},t) = Te^{i(\kappa_a x - \omega t)}\\
\label{eq_comp2}
&p^{(r)}(\bs{x},t) = -i\rho_a v_a \omega T e^{i(\kappa_a x - \omega t)}
\end{align}
where $T$ is the transmission coefficient, i.e. the fraction of the incident wave's amplitude that is transmitted through the panel. To find $R$ and $T$, a simple 2D FE discretization of $4\times4$ quadrilateral elements with 4 gaussian integration points over a portion of the panel is performed. The resulting system is solved in the frequency domain for each frequency in the desired range, in this case from 0 to 3000 Hz. Periodic boundary conditions are applied on the top and bottom boundary nodes (to simulate the infinite extension of the panel in the vertical direction) and the compatibility conditions given by equations \eqref{eq_comp1} to \eqref{eq_comp2} are applied on the left and right boundary nodes. The resulting system can be expressed in matrix form as
\begin{equation}
\label{eq_system_macro_TL}
	(\mathbb{K} - i \omega \mu \mathbb{C} - \omega^2 \mathbb{M}) \hat{\bs{U}}(R,T) = \hat{\bs{F}}(R,T).
\end{equation}
Since both $\hat{\bs{U}}$ and $\hat{\bs{F}}$ are functions of the reflection and transmission coefficients (as a result of applying the compatibility conditions), equation \eqref{eq_system_macro_TL} can be reduced to a system with only $R$ and $T$ as unknowns (details on the derivation of this system are given in \ref{app_RT_system}). Thus, for a given frequency $\omega$ and viscosity $\mu$, once the transmission coefficient $T$ is solved, the transmission loss is computed using the expression
\begin{equation}
	\text{TL}(\omega,\mu) = -20 \log_{10} |T(\omega,\mu)| \ \ [\text{dB}].
\end{equation}

\begin{figure*}
	\centering
	\includegraphics{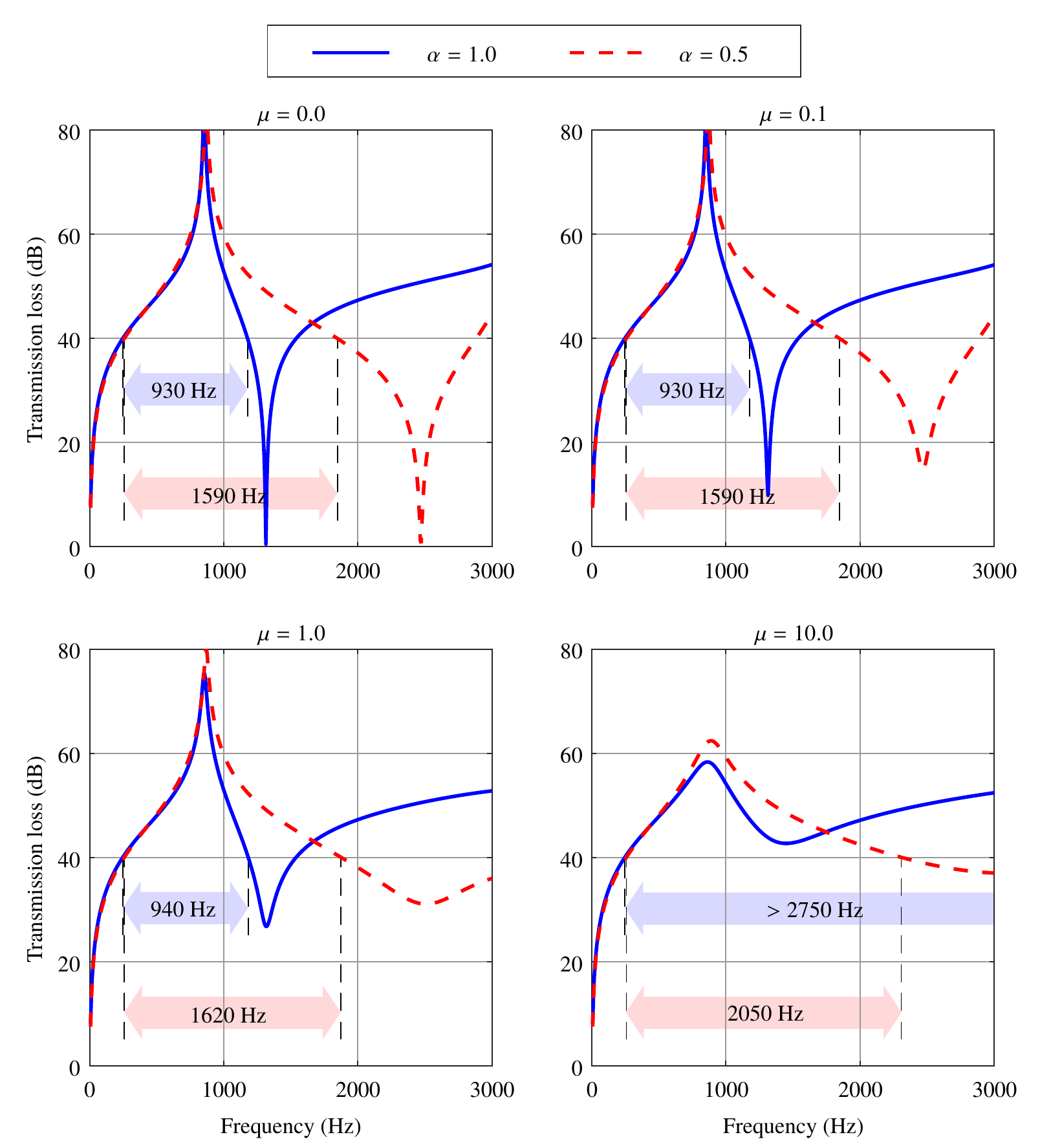}
	\caption{Transmission loss (TL) for the topologies resulting from the frequency fitting optimization (solid blue line) and the frequency fitting coupled with the bandgap maximization (red dashed line). The light-shaded arrows indicate the bandwidth of the effective attenuation bands corresponding to the first uninterrupted attenuations $>40$ dB. The units for the viscosity parameter $\mu$ are Pa$\cdot$s.}
	\label{fig8}
\end{figure*}

The results are shown in Figure \ref{fig8} where the transmission loss has been computed for a panel with thickness of 1 cm (equivalent to a single RVE) and for both the topologies without and with the bandgap maximized. One can observe that the effect of the bandgap maximization ($\alpha=0.5$) translates also into a larger frequency range of effective attenuation. For instance, for attenuation levels above 40 dB, the effective bands in both cases start around 250 Hz, but they extend to 1180 Hz in the first case and to 1840 in the second (660 Hz increase).


It can also be observed that the viscosity has the beneficial effect of smoothing the undesired inverted resonance peaks but at the same time it also dampens the attenuation peaks caused by the local resonance phenomena occurring at the microscale. Depending on the desired attenuation performance, one can take advantage of the viscoelastic properties of the coating component to slightly increase the effective attenuation band or even bypass the undesired inverted resonance peak while keeping a continuous effective attenuation band with a good level of attenuation in the low frequency range of interest (as it can be seen for the case of $\mu=10$ Pa$\cdot$s, where viscous effects are more relevant).

\section{Concluding remarks}
\label{sec_conclusions}
While the methodology presented has been used in the study of flat panels under plane acoustic pressure waves, it can be suitably adapted to more complex cases still satisfying the low-frequency range restriction. On the one hand, the objective function of the topology optimization algorithm can be adapted to tackle one or multiple frequencies depending on the user interest. On the other hand, the ability of the homogenization procedure to provide a set of effective \textit{constitutive} properties makes it possible to easily study the behaviour of complex geometries in the macroscale under several sets of boundary conditions and external actions, including, for instance, the characterization of transient states or multiple layer configurations, so long as they satisfy the hypotheses considered. This is in contrast to other homogenization approaches, such as those based on the Bloch-Floquet theory, which have the ability to study the effective behaviour of RVEs without restrictions in terms of frequencies, but that are limited to predetermined macroscopic configurations, namely, infinitely periodic structures, in the case of the Bloch-Floquet theory, that cannot account for the introduction of other more realistic sets of boundary conditions.  

In this regard, the design procedure presented in this work offers a powerful tool for the design of LRAM acoustic insulation devices tackling the low frequency range. By combining the optimization algorithm for designing the RVE topology with the homogenization method for characterising the material's performance, one can easily assess the transmission loss of a panel for a broad range of frequencies. The studies carried out show the influence of the RVE topology in achieving certain properties in the final design such as, for instance, an increased bandgap size for a given frequency range with the same amount of material, and thus the same mass. It is worth noting that while the target frequencies and choice of materials greatly determine the size of the RVE, its topology optimization can be used to reduce the effective density making the resulting panel suitable for lightweight applications. On the other hand, it has been observed that the presence of highly viscoelastic materials in the design generally affects negatively the local resonance performance of the RVE. However, in certain viscosity ranges, one can take advantage of the resulting damping effects to smooth undesired resonance peaks in the macroscale and even join attenuation bands, specially in higher frequency ranges, while still maintaining good attenuation levels in the low-frequency range of interest where the local resonance of the RVE takes place.

\section*{Acknowledgements}
This research has received funding from the European Research Council (ERC) under the European Union's Horizon 2020 research and innovation programme (Proof of Concept Grant agreement nº 779611) through the project ``Computational catalog of multiscale materials: a plugin library for industrial finite element codes'' (CATALOG). The authors also acknowledge the funding received from the Spanish Ministry of Economy and Competitiveness through the research grant DPI2017-85521-P for the project ``Computational design of Acoustic and Mechanical Metamaterials'' (METAMAT).

\appendix

\section{Introduction of viscoelastic effects in the homogenization framework}
\label{annex_introduction_viscoelastic}
Considering equation \eqref{eq_micro_stress_viscoelastic} for the stress definition in the microscale, causes a damping matrix, $\mathbb{C}_\mu$, to arise from the Finite Element discretization of the RVE system, in addition to the standard mass and stiffness matrices, $\mathbb{M}_\mu$ and $\mathbb{K}_\mu$. This damping matrix appears only in the elements where viscoelastic effects are considered, in which it is defined as
\begin{equation}
	\mathbb{C}_\mu^{(e)} = \avg{\bs{B}_\mu^{(e)\text{T}} : \bms{\eta}_\mu^{(e)} : \bs{B}_\mu^{(e)}}_{\Omega_\mu^{(e)}}, 
\end{equation}
where $\bs{B}_\mu^{(e)}$ gives the derivatives of the shape functions for the type of elements considered, and $\bms{\eta}_\mu^{(e)}$ is the element's viscous tensor, which can be expressed, following Voigt's notation, in terms of the associated material's viscosity, $\mu^{(e)}$, as
\begin{equation}
	\bms{\eta}_\mu^{(e)} = \mu^{(e)}
	\begin{bmatrix}
	 4/3 & -2/3 & -2/3 & 0 & 0 & 0 \\
	-2/3 &  4/3 & -2/3 & 0 & 0 & 0 \\
	-2/3 & -2/3 &  4/3 & 0 & 0 & 0 \\
	   0 &    0 &    0 & 1 & 0 & 0 \\
	   0 &    0 &    0 & 0 & 1 & 0 \\
	   0 &    0 &    0 & 0 & 0 & 1
	\end{bmatrix}.
\end{equation}

After a standard assembly process, denoted here with the big A symbol, one can obtain the global damping matrix
\begin{equation}
	\mathbb{C}_\mu = \bigA_e \mathbb{C}_\mu^{(e)},
\end{equation}
and the extended RVE dynamic system results
\begin{align}
\label{eq_micro_extended_system_damping}
	\begin{bmatrix}
		\mathbb{M}_\mu & \bs{0} & \bs{0} \\
		\bs{0} & \bs{0} & \bs{0} \\
		\bs{0} & \bs{0} & \bs{0}
	\end{bmatrix}
	\begin{bmatrix}
		\ddot{\hat{\bs{u}}}_\mu \\
		\ddot{\bms{\beta}} \\
		\ddot{\bms{\lambda}}
	\end{bmatrix} + 
	\begin{bmatrix}
		\mathbb{C}_\mu & \bs{0} & \bs{0} \\
		\bs{0} & \bs{0} & \bs{0} \\
		\bs{0} & \bs{0} & \bs{0}
	\end{bmatrix}
	\begin{bmatrix}
		\dot{\hat{\bs{u}}}_\mu \\
		\dot{\bms{\beta}} \\
		\dot{\bms{\lambda}}
	\end{bmatrix} + \begin{bmatrix}
		\mathbb{K}_\mu & -\mathbb{N}_\mu^\text{T} & -\mathbb{B}_\mu^\text{T} \\
		-\mathbb{N}_\mu & \bs{0} & \bs{0} \\
		-\mathbb{B}_\mu & \bs{0} & \bs{0}
	\end{bmatrix}
	\begin{bmatrix}
		\hat{\bs{u}}_\mu \\
		\bms{\beta} \\
		\bms{\lambda}
	\end{bmatrix} = 
	\begin{bmatrix}
		\bs{0} \\
		-\bs{u} \\
		-\bms{\varepsilon}
	\end{bmatrix}. 
\end{align}
Note also that matrices denoted by $\mathbb{N}_\mu$ and $\mathbb{B}_\mu$ appear in system \eqref{eq_micro_extended_system_damping}. These come from the discretization of the minimal kinematic restrictions, as explained in detail in \citet{Roca2018}. As a reminder,
\begin{align}
\label{eq_micro_N}
	&\mathbb{N}_\mu = \bigA_e \avg{\bs{N}_\mu^{(e)}}_{\Omega_\mu^{(e)}}, \quad \text{so} \quad \mathbb{N}_\mu \hat{\bs{u}}_\mu \equiv \avg{\bs{u}_\mu}_{\Omega_\mu} = \bs{u},  \\
\label{eq_micro_B}
	&\mathbb{B}_\mu = \bigA_e \avg{\bs{B}_\mu^{(e)}}_{\Omega_\mu^{(e)}},  \quad \text{so} \quad \mathbb{B}_\mu \hat{\bs{u}}_\mu \equiv \avg{\nabla_\bs{y}^\text{S}\bs{u}_\mu}_{\Omega_\mu} = \bms{\varepsilon},
\end{align}
where $\bs{N}_\mu^{(e)}$ corresponds to the shape functions for the type of elements considered.

Since the hypotheses for the split of the system \eqref{eq_micro_extended_system_damping} into its quasi-static and inertial components still hold, despite the introduction of viscoelastic effects in the framework, one obtains:
\begin{itemize}
	\item[a)] \textit{Quasi-static system}
	\begin{equation}
	\label{eq_micro_quasi-static_system_damping}
		\begin{bmatrix}
			\mathbb{C}_\mu & \bs{0} \\
			\bs{0} & \bs{0}
		\end{bmatrix}
		\begin{bmatrix}
			\dot{\hat{\bs{u}}}_\mu^{(1)} \\
			\dot{\bms{\lambda}}^{(1)}
		\end{bmatrix} + 
		\begin{bmatrix}
			\mathbb{K}_\mu & -\mathbb{B}_\mu^\text{T} \\
			-\mathbb{B}_\mu & \bs{0}
		\end{bmatrix}
		\begin{bmatrix}
			\hat{\bs{u}}_\mu^{(1)} \\
			\bms{\lambda}^{(1)}
		\end{bmatrix} =
		\begin{bmatrix}
			\bs{0} \\
			-\bms{\varepsilon}
		\end{bmatrix}.
	\end{equation}
	For the sake of generality, let us assume that kinematic conditions are imposed directly on the system through
	\begin{equation}
		\bs{u}_\mu^{(1)} = \bms{\varepsilon} \cdot \Delta\bs{y} + \tilde{\bs{u}}_\mu^{(1)},
	\end{equation}
	where $\Delta\bs{y} = \bs{y} - \bs{y}^{(0)}$, and $\bs{y}^{(0)}$ is the centroid of the RVE ($\bs{y}^{(0)} = \avg{\bs{y}}_{\Omega_\mu}$). In discretized form, this is
	\begin{equation}
	\label{eq_micro_conditions_strain}
		\hat{\bs{u}}_\mu^{(1)} = 
		\begin{bmatrix}
			\mathbb{Y} & \mathbb{P} _\mu
		\end{bmatrix} 
		\begin{bmatrix}
			\bms{\varepsilon} \\
			\hat{\bs{u}}_\mu^{*(1)}
		\end{bmatrix},
	\end{equation} 
	with 
	\begin{equation}
		\mathbb{Y} = 
		\begin{bmatrix}
			\vdots \\
			\begin{matrix}
				\Delta\hat{y}_1^{(j)} & 0 & 0 & 0 & \Delta\hat{y}_3^{(j)}/2 & \Delta\hat{y}_2^{(j)}/2 \\
				0 & \Delta\hat{y}_2^{(j)} & 0 & \Delta\hat{y}_3^{(j)}/2 & 0 & \Delta\hat{y}_1^{(j)}/2 \\
				0 & 0 & \Delta\hat{y}_3^{(j)} & \Delta\hat{y}_2^{(j)}/2 & \Delta\hat{y}_3^{(j)}/2 & 0 \\
			\end{matrix}\\
			\vdots
		\end{bmatrix}.
	\end{equation}
	and $\mathbb{P}_\mu$ being a matrix that imposes the desired boundary conditions on the discretized microfluctuation field, $\hat{\tilde{\bs{u}}}_\mu^{(1)}$. Typically, for the kind of problems tackled, periodic boundary conditions along with prescription of certain degrees of freedom to prevent rigid body motions offer good results. In such cases,
	\begin{equation}
	\label{eq_micro_matrix_prescription}
		\begin{bmatrix}
			\hat{\tilde{\bs{u}}}_\mu^{(1)0} \\
			\hat{\tilde{\bs{u}}}_\mu^{(1)i} \\
			\hat{\tilde{\bs{u}}}_\mu^{(1)+} \\
			\hat{\tilde{\bs{u}}}_\mu^{(1)-}
		\end{bmatrix} =
		\underbrace{\begin{bmatrix}
			\bs{0} & \bs{0} \\
			\bs{I} & \bs{0} \\
			\bs{0} & \bs{I} \\
			\bs{0} & \bs{I}
		\end{bmatrix}}_{\mathbb{P}_\mu}
		\begin{bmatrix}
		\hat{\bs{u}}_\mu^{*(1)i} \\
		\hat{\bs{u}}_\mu^{*(1)+}
		\end{bmatrix},
	\end{equation}
	where superscripts 0, $i$, $+$ and $-$ are used to refer to prescribed, internal, and periodic boundaries degrees of freedom, respectively.
	
	Note that, from the definition of matrix $\mathbb{B}_\mu$ (see equation \eqref{eq_micro_B}), it can be seen that $\mathbb{B}_\mu \mathbb{Y} \equiv \avg{\nabla_\bs{y}^\text{S}\Delta\bs{y}}_{\Omega_\mu} = \bs{I}$. Thus, by premultiplying the first equation in system \eqref{eq_micro_quasi-static_system_damping} by $\mathbb{Y}^\text{T}$ and introducing expression \eqref{eq_micro_conditions_strain} allows us to find, after some algebraic manipulation,
	\begin{equation}
	\label{eq_lambda_1_viscoelastic}
		\bms{\lambda}^{(1)} = \mathbb{Y}^\text{T} \mathbb{C}_\mu (\mathbb{Y} \dot{\bms{\varepsilon}} + \mathbb{P}_\mu \dot{\hat{\bs{u}}}_\mu^{*(1)}) + \mathbb{Y}^\text{T} \mathbb{K}_\mu (\mathbb{Y} \bms{\varepsilon} + \mathbb{P}_\mu \hat{\bs{u}}_\mu^{*(1)}).
	\end{equation}
	
	Assuming the kinematic restriction imposed is compatible with that associated to the Lagrange multiplier, projecting the system into a set satisfying the microfluctuation field boundary conditions, i.e. premultiplying the first equation in system \eqref{eq_micro_quasi-static_system_damping} by $\mathbb{P}_\mu^\text{T}$, results in	
	\begin{equation}
	\label{eq_micro_quasi-static_system_damping_prescribed_1}
		\mathbb{P}_\mu^\text{T} \mathbb{C}_\mu \mathbb{P}_\mu \dot{\hat{\bs{u}}}_\mu^{*(1)} + \mathbb{P}_\mu^\text{T} \mathbb{K}_\mu \mathbb{P}_\mu \hat{\bs{u}}_\mu^{*(1)} = -\mathbb{P}_\mu^\text{T} \mathbb{C}_\mu \mathbb{Y} \dot{\bms{\varepsilon}} - \mathbb{P}_\mu^\text{T} \mathbb{K}_\mu \mathbb{Y} \bms{\varepsilon}.
	\end{equation}
	Taking the time derivative of equation \eqref{eq_micro_quasi-static_system_damping_prescribed_1} gives 
	\begin{equation}
	\label{eq_micro_quasi-static_system_damping_prescribed_2}
		\mathbb{P}_\mu^\text{T} \mathbb{K}_\mu \mathbb{P}_\mu \dot{\hat{\bs{u}}}_\mu^{*(1)} = -\mathbb{P}_\mu^\text{T} \mathbb{K}_\mu \mathbb{Y} \dot{\bms{\varepsilon}},
	\end{equation}
	where the hypothesis for the quasi-static system, $\ddot{\bms{\varepsilon}} \approx \bs{0}$ (which makes $\ddot{\hat{\bs{u}}}_\mu^{*(1)} \approx \bs{0}$), has been considered.
	
	The solution for $\dot{\hat{\bs{u}}}_\mu^{*(1)}$ obtained from equation \eqref{eq_micro_quasi-static_system_damping_prescribed_2} in terms of the macroscopic strain rate, $\dot{\bms{\varepsilon}}$, can be introduced into equation \eqref{eq_micro_quasi-static_system_damping_prescribed_1} to compute the solution field 
	\begin{equation}
	\label{eq_micro_quasi-static_system_solution}
		\hat{\bs{u}}_\mu^{*(1)} = -(\mathbb{P}_\mu^\text{T} \mathbb{K}_\mu \mathbb{P})_\mu^{-1}\mathbb{P}_\mu^\text{T}(\mathbb{K}_\mu \mathbb{Y} \bms{\varepsilon} + \mathbb{C}_\mu \widetilde{\mathbb{Y}} \dot{\bms{\varepsilon}}),
	\end{equation}
	with
	\begin{equation}
		\widetilde{\mathbb{Y}} = (\bs{I} - \mathbb{P}_\mu (\mathbb{P}_\mu^\text{T} \mathbb{K}_\mu \mathbb{P}_\mu)^{-1} \mathbb{P}_\mu^\text{T} \mathbb{K}_\mu)\mathbb{Y}.
	\end{equation}
	Finally, substituting expression \eqref{eq_micro_quasi-static_system_solution} into equation \eqref{eq_lambda_1_viscoelastic}, yields
	\begin{equation}
	\label{eq_lambda_1_viscoelastic_final}
		\bms{\lambda}^{(1)} = \underbrace{\mathbb{Y}^\text{T}\mathbb{K}_\mu\widetilde{\mathbb{Y}}}_{\bs{C}^\text{eff}} : \bms{\varepsilon} + \underbrace{\widetilde{\mathbb{Y}}^\text{T}\mathbb{C}_\mu\widetilde{\mathbb{Y}}}_{\bms{\eta}^\text{eff}} : \dot{\bms{\varepsilon}}.
	\end{equation}	
	Note that $\bs{C}^\text{eff}$ in equation \eqref{eq_lambda_1_viscoelastic_final} assumes the role of an effective constitutive tensor and it is the same than the one obtained in \citet{Roca2018}, without considering viscoelastic effects. In fact, the quasi-static influence of viscoelasticity in the model is accounted by the new term $\bms{\eta}^\text{eff}$, which acts as a sort of effective viscous tensor relating the macroscopic stress with strain rates. 
	
	\item[b)] \textit{Inertial system}
	\begin{align}
	\label{eq_micro_inertial_system_damping}
		\begin{bmatrix}
			\mathbb{M}_\mu & \bs{0} \\
			\bs{0} & \bs{0}
		\end{bmatrix}
		\begin{bmatrix}
			\ddot{\hat{\bs{u}}}_\mu^{(2)} \\
			\ddot{\bms{\beta}}^{(2)}
		\end{bmatrix} + 
		\begin{bmatrix}
			\mathbb{C}_\mu & \bs{0} \\
			\bs{0} & \bs{0}
		\end{bmatrix}
		\begin{bmatrix}
			\dot{\hat{\bs{u}}}_\mu^{(2)} \\
			\dot{\bms{\beta}}^{(2)}
		\end{bmatrix} + \begin{bmatrix}
			\mathbb{K}_\mu & -\mathbb{N}_\mu^\text{T} \\
			-\mathbb{N}_\mu & \bs{0}
		\end{bmatrix}
		\begin{bmatrix}
			\hat{\bs{u}}_\mu^{(2)} \\
			\bms{\beta}^{(2)}
		\end{bmatrix} =
		\begin{bmatrix}
			\bs{0} \\
			-\bs{u}
		\end{bmatrix}.
	\end{align}
	
	Following a similar procedure than in the quasi-static case, in order to maintain the generality of the formulation, the specific kinematic conditions considered here will also be introduced directly in the system \eqref{eq_micro_inertial_system_damping} by taking
	\begin{equation}
		\bs{u}_\mu^{(2)} = \bs{u} + \tilde{\bs{u}}_\mu^{(2)},
	\end{equation}
	which can be expressed in discretized form as
	\begin{equation}
	\label{eq_micro_inertial_prescribed}
		\hat{\bs{u}}_\mu^{(2)} = 
		\begin{bmatrix}
			\mathbb{I} & \mathbb{P}_\mu 
		\end{bmatrix} 
		\begin{bmatrix}
			\bs{u} \\
			\hat{\bs{u}}_\mu^{*(2)}
		\end{bmatrix},
	\end{equation} 
	with
	\begin{equation}
		\mathbb{I} = 
		\begin{bmatrix}
			\vdots \\
			\begin{matrix}
				1 & 0 & 0 \\
				0 & 1 & 0 \\
				0 & 0 & 1 \\
			\end{matrix}\\
			\vdots
		\end{bmatrix},
	\end{equation}
	and $\mathbb{P}$ being, again, a matrix imposing the desired boundary conditions over the microfluctuation field (see, for instance, eq. \eqref{eq_micro_matrix_prescription}).
	
	In this case, it is easy to verify, from the matrix $\mathbb{N}_\mu$ definition in equation \eqref{eq_micro_N}, that $\mathbb{N}_\mu\mathbb{I} = \bs{I}$. 	Also note that, since $\mathbb{I}$ is a rigid body translation mode, it belongs to both the kernels of $\mathbb{K}_\mu$ and $\mathbb{C}_\mu$ (i.e. $\mathbb{K}_\mu \mathbb{I} = \bs{0}$ and $\mathbb{C}_\mu \mathbb{I} = \bs{0}$). These properties allow us to obtain, after premultiplying the first equation in system \eqref{eq_micro_inertial_system_damping} by $\mathbb{I}^\text{T}$ and some algebraic manipulation,
	\begin{equation}
	\label{eq_beta_2}
		\bms{\beta}^{(2)} = \underbrace{\mathbb{I}^\text{T} \mathbb{M}_\mu \mathbb{I}}_{\bar{\rho}\bs{I} = \avg{\rho_\mu}_{\Omega_\mu}\bs{I}} \ddot{\bs{u}} + \underbrace{\mathbb{I}^\text{T} \mathbb{M}_\mu \mathbb{P}_\mu}_{\mathbb{D}} \ddot{\hat{\bs{u}}}_\mu^{*(2)}.
	\end{equation}
	 
	Again, projecting the system \eqref{eq_micro_inertial_system_damping} into a set satisfying the imposed kinematic restrictions on the microfluctuation field through $\mathbb{P}_\mu$ (nullifying the effect of the Lagrange multiplier), allows us to obtain
	\begin{align}
	\label{eq_micro_inertial_system_damping_prescribed}
		\underbrace{\mathbb{P}_\mu^\text{T} \mathbb{M}_\mu \mathbb{P}_\mu}_{\mathbb{M}_\mu^{*}} \ddot{\hat{\bs{u}}}_\mu^{*(2)} + \underbrace{\mathbb{P}_\mu^\text{T} \mathbb{C}_\mu \mathbb{P}_\mu}_{\mathbb{C}_\mu^{*}} \dot{\hat{\bs{u}}}_\mu^{*(2)}+ \underbrace{\mathbb{P}_\mu^\text{T} \mathbb{K}_\mu \mathbb{P}_\mu}_{\mathbb{K}_\mu^{*}} \hat{\bs{u}}_\mu^{*(2)} = -\underbrace{\mathbb{P}_\mu^\text{T} \mathbb{M}_\mu \mathbb{I}}_{\mathbb{D}^\text{T}} \ddot{\bs{u}}. 
	\end{align}
	In order to make the model computationally efficient, a model order reduction is performed by projecting the solution field $\hat{\bs{u}}_\mu^{*(2)}$ onto the space spanned by the eigenmodes of the \textit{undamped} system \eqref{eq_micro_inertial_system_damping_prescribed}
	\begin{equation}
		(\mathbb{K}_\mu^{*} - \lambda_\mu^{(k)} \mathbb{M}_\mu^{*})\hat{\bms{\phi}}_\mu^{(k)} = \bs{0},
	\end{equation}
	where $\lambda_\mu^{(k)} = (\omega_\mu^{(k)})^2$ here refer to the squared natural frequencies and $\hat{\bms{\phi}}_\mu^{(k)}$ are the associated mass-normalized vibration modes (i.e. $\hat{\bms{\phi}}_\mu^{(k)\text{T}}\mathbb{M}_\mu^{*}\hat{\bms{\phi}}_\mu^{(k)} = 1$ and $\hat{\bms{\phi}}_\mu^{(k)\text{T}}\mathbb{K}_\mu^{*}\hat{\bms{\phi}}_\mu^{(k)} = (\omega_\mu^{(k)})^2$)\new{\footnote{\new{Code functionalities allowing the computation of only the smallest eigenvalues in magnitude and their associated eigenvectors (which are potential candidates to become relevant modes in the frequency range of interest) have been used, this translating into a reduction of the computational cost of the modal problem evaluation.}}}. Now, defining $\bms{\Omega}_\mu$ as the diagonal matrix with only the natural frequencies in the range of interest, $\bms{\Phi}_\mu$ as a matrix that contains their associated mass-normalized eigenmodes, and $\bs{q}_\mu$ as the column vector with their corresponding modal amplitudes, one can express 
	\begin{equation}
		\hat{\bs{u}}_\mu^{*(2)} = \bms{\Phi}_\mu\bs{q}_\mu,
	\end{equation} 
	and equations \eqref{eq_beta_2} and \eqref{eq_micro_inertial_system_damping_prescribed} become
	\begin{align}
		&\bms{\beta}^{(2)} = \bar{\rho} \ddot{\bs{u}} + \underbrace{\mathbb{D}\bms{\Phi}_\mu}_{\mathbb{Q}} \ddot{\bs{q}}_\mu, \\
		&\ddot{\bs{q}}_\mu + \underbrace{\bms{\Phi}_\mu^\text{T}\mathbb{C}_\mu^{*}\bms{\Phi}_\mu}_{\bms{\Omega}_\mu^\text{D}} \dot{\bs{q}}_\mu + \bms{\Omega}_\mu^2 \bs{q}_\mu = -\underbrace{\bms{\Phi}_\mu^\text{T}\mathbb{D}^\text{T}}_{\mathbb{Q}^\text{T}} \ddot{\bs{u}}.
	\end{align}
\end{itemize}

\section{Sensitivity of the LRAM topology optimization cost function}
\label{app_sensitivity_cost_fcn}
For any functional of the form
\begin{equation}
\Pi_t(\chi(\phi)) = \int_{\Omega_\mu} \pi(\chi(\phi(\bs{y},t))) d\Omega
\end{equation}
the VTD at any point $\hat{\bs{y}}$ in the domain is given by 
\begin{equation}
\label{eq_VTD}
\dfrac{\delta \Pi}{\delta \chi}(\hat{\bs{y}}) = \int_{\Omega_\mu} \dfrac{\partial \pi(\chi(\phi(\bs{y},t)))}{\partial \chi} \delta_{\hat{\bs{y}}} d\Omega = \left.\dfrac{\partial \pi(\bs{y},t)}{\partial \chi}\right|_{\bs{y}=\hat{\bs{y}}}
\end{equation}
where $\delta_{\hat{\bs{y}}}$ is the point-Dirac's delta shifted to point $\hat{\bs{y}}$ fulfilling
\begin{equation}
\int_{\Omega_\mu} f(\bs{y}) \delta_{\hat{\bs{y}}} d\Omega = f(\hat{\bs{y}}).
\end{equation}
On the other hand, the evolution of the cost function can be computed, accounting for equation \eqref{eq_HJac_rate}, as
\begin{align}
\label{eq_dot_pi}
\dot{\Pi} & {} = \int_{\Omega_\mu} \dfrac{\partial \pi}{\partial \chi} \dfrac{\partial \chi}{\partial \phi} \dot{\phi} d\Omega = -C_1 \int_{\Omega_\mu} \dfrac{\partial \pi}{\partial \chi} \dfrac{\partial \chi}{\partial \phi} \underbrace{\dfrac{\delta\Pi}{\delta\chi}(\bs{y})}_{\displaystyle{\frac{\partial \pi}{\partial \chi}(\bs{y},t)}} d\Omega = -C_1 \int_{\Omega_\mu} \dfrac{\partial \chi}{\partial \phi} \left(\dfrac{\partial \pi}{\partial \chi}\right)^2  d\Omega,
\end{align}
where equation \eqref{eq_VTD} has been considered. From equation \eqref{eq_chi}, it can be proven that
\begin{equation}
\label{eq_der_chi}
\dfrac{\partial \chi}{\partial \phi} = \left\| \nabla\phi \right\|^{-1} \delta_{\Gamma_\phi}
\end{equation}
where $\delta_{\Gamma_\phi}$ stands for the line/surface-Dirac's delta shifted to the zero level-set of $\phi$ in $\Omega_\mu$, i.e. $\Gamma_\phi:=\{\bs{y}\in\Omega_\mu \ | \ \phi(\bs{y}) = 0\}$, thus fulfilling
\begin{equation}
\label{eq_gamma_psi}
\int_{\Omega_\mu} f(\bs{y}) \delta_{\Gamma_\phi} d\Omega = \int_{\Gamma_\phi} f(\bs{y}) d\Gamma.
\end{equation}
Replacing equations \eqref{eq_der_chi} and \eqref{eq_gamma_psi} into equation \eqref{eq_dot_pi} yields
\begin{align}
\label{eq_decrease}
\dot{\Pi} = -C_1 \int_{\Gamma_\phi} \left\| \nabla\phi \right\|^{-1} \left(\dfrac{\partial \pi}{\partial \chi}\right)^2  d\Gamma \leq 0.
\end{align}
Equation \eqref{eq_decrease} proofs the descending character of the cost function $\Pi$ along time/iteration evolution.

The sensitivity of the cost function \eqref{eq_objective_fcn} will be evaluated using the variational topological derivative (VTD). To do so, first the chain rule will be applied to equation \eqref{eq_objective_fcn} to obtain
\begin{align}
&\dfrac{\delta\Pi}{\delta\chi} = \dfrac{4\alpha f}{\lambda_\mu^{*(1)} \ln \bar{\lambda}_\mu^{*}} \left( \dfrac{\ln \bar{\lambda}_\mu^{*}}{\ln \lambda_\mu^{*(1)} + \ln \bar{\lambda}_\mu^{*}} \right)^2 \dfrac{\delta\lambda_\mu^{*(1)}}{\delta\chi} + \dfrac{2(1-\alpha) g}{\lambda_\mu^{*(1)} \ln \lambda_\mu^{(1)}} \left( \dfrac{\delta\lambda_\mu^{*(1)}}{\delta\chi} - g \dfrac{\lambda_\mu^{*(1)}}{\lambda_\mu^{(1)}} \dfrac{\delta\lambda_\mu^{(1)}}{\delta\chi} \right).
\end{align}
Now, in order to compute $\delta\lambda_\mu^{*(1)}/\delta\chi$ and $\delta\lambda_\mu^{(1)}/\delta\chi$ let us take the derivative of the state-equations \eqref{eq_modal_restricted} and \eqref{eq_modal_unrestricted}, which gives
\begin{align}
\label{eq_modal_restricted_deriv}
&\mathbb{M}_\mu^\ast \hat{\bms{\phi}}_\mu^{*(1)} \dfrac{\delta\lambda_\mu^{*(1)}}{\delta\chi} = \left( \dfrac{\delta\mathbb{K}_\mu^\ast}{\delta\chi} - \lambda_\mu^{*(1)} \dfrac{\delta\mathbb{M}_\mu^\ast}{\delta\chi} \right) \hat{\bms{\phi}}_\mu^{*(1)} + ( \mathbb{K}_\mu^\ast - \lambda_\mu^{*(1)} \mathbb{M}_\mu^\ast ) \dfrac{\delta\hat{\bms{\phi}}_\mu^{*(1)}}{\delta\chi}, \\
\label{eq_modal_unrestricted_deriv}
&\mathbb{M}_\mu \hat{\bms{\phi}}_\mu^{(1)} \dfrac{\delta\lambda_\mu^{(1)}}{\delta\chi} = \left( \dfrac{\delta\mathbb{K}_\mu}{\delta\chi} - \lambda_\mu^{(1)} \dfrac{\delta\mathbb{M}_\mu}{\delta\chi} \right) \hat{\bms{\phi}}_\mu^{(1)} + ( \mathbb{K}_\mu - \lambda_\mu^{(1)} \mathbb{M}_\mu ) \dfrac{\delta\hat{\bms{\phi}}_\mu^{(1)}}{\delta\chi}.
\end{align}
Pre-multiplying equations \eqref{eq_modal_restricted_deriv} and \eqref{eq_modal_unrestricted_deriv} by $\hat{\bms{\phi}}_\mu^{*(1)\text{T}}$ and $\hat{\bms{\phi}}_\mu^{(1)\text{T}}$, respectively, allows us to obtain
\begin{align}
&\dfrac{\delta\lambda_\mu^{*(1)}}{\delta\chi} = \hat{\bms{\phi}}_\mu^{*(1)\text{T}}\left( \dfrac{\delta\mathbb{K}_\mu^\ast}{\delta\chi} - \lambda_\mu^{*(1)} \dfrac{\delta\mathbb{M}_\mu^\ast}{\delta\chi} \right) \hat{\bms{\phi}}_\mu^{*(1)},&\\
&\dfrac{\delta\lambda_\mu^{(1)}}{\delta\chi} = \hat{\bms{\phi}}_\mu^{(1)\text{T}}\left( \dfrac{\delta\mathbb{K}_\mu}{\delta\chi} - \lambda_\mu^{(1)} \dfrac{\delta\mathbb{M}_\mu}{\delta\chi} \right) \hat{\bms{\phi}}_\mu^{(1)}.&
\end{align}
Notice that the fact the vibration modes in each system are mass-normalized has been used. See also how the term multiplying the derivatives of the vibration modes in each system vanishes due to equations \eqref{eq_modal_restricted} and \eqref{eq_modal_unrestricted}. According to the VTD definition, one finds
\begin{align}
&\hat{\bms{\phi}}_\mu^{(1)\text{T}} \dfrac{\delta\mathbb{K}_\mu}{\delta\chi} (\hat{\bs{y}})  \hat{\bms{\phi}}_\mu^{(1)} \equiv \nabla_\bs{y}^\text{S} \bms{\phi}_\mu^{(1)}(\hat{\bs{y}}) : \dfrac{\partial\bs{C}_\mu(\chi(\hat{\bs{y}}))}{\partial\chi} : \nabla_\bs{y}^\text{S} \bms{\phi}_\mu^{(1)}(\hat{\bs{y}}), \\
&\hat{\bms{\phi}}_\mu^{(1)\text{T}} \dfrac{\delta\mathbb{M}_\mu}{\delta\chi} (\hat{\bs{y}}) \hat{\bms{\phi}}_\mu^{(1)} \equiv \dfrac{\partial\rho_\mu(\chi(\hat{\bs{y}}))}{\partial\chi} \|\bms{\phi}_\mu^{(1)}(\hat{\bs{y}})\|^2,
\end{align}
where $\partial\bs{C}_\mu(\chi(\hat{\bs{y}}))/\partial \chi$ and $\partial\rho_\mu(\chi(\hat{\bs{y}}))/\partial\chi$ are regular function derivatives of the constitutive tensor and density distribution on the design domain evaluated at point $\hat{\bs{y}}$. In the context of linear elastic isotropic behaviour of the material phases, let us now consider $K_\mu^+$, $G_\mu^+$ and $\rho_\mu^+$ the bulk modulus, shear modulus and density of the \textit{dense material} region $\Omega_\mu^+$ (inclusions) and $K_\mu^-$, $G_\mu^-$ and $\rho_\mu^-$ the bulk modulus, shear modulus and density of the \textit{soft material} region $\Omega_\mu^-$ (void/coating), such that
\begin{align}
&\bs{C}_\mu(\chi) = \hat{K}_\mu(\chi) \bs{I} \otimes \bs{I} + 2 \hat{G}_\mu(\chi) \bs{I}^\text{dev} \\
&\rho_\mu(\chi) = \hat{\rho_\mu}(\chi),
\end{align}
where $\hat{K}_\mu$, $\hat{G}_\mu$, $\hat{\rho_\mu}$ are interpolation functions of the form 
\begin{equation}
\label{eq_interp}
	\hat{h}(\chi) = \left[ \chi (h^+)^\frac{1}{n} + (1-\chi)(h^-)^\frac{1}{n} \right]^n.
\end{equation}
Note that for $n>0$ (typically a value of 2 is chosen), equation \eqref{eq_interp} returns $h^+$ for dense material regions ($\chi=1$) and $h^-$ for soft material regions ($\chi=0$). Eventually, one can compute
\begin{align}
&\dfrac{\partial\bs{C}_\mu(\chi)}{\partial\chi} = n\left(\hat{K}_\mu(\chi)\right)^\frac{n-1}{n} \left((K_\mu^+)^\frac{1}{n} - (K_\mu^-)^\frac{1}{n}\right) \bs{I} \otimes \bs{I} + 2 n\left(\hat{G}_\mu(\chi)\right)^\frac{n-1}{n} \left((G_\mu^+)^\frac{1}{n} - (G_\mu^-)^\frac{1}{n}\right) \bs{I}^\text{dev},  \\
&\dfrac{\partial\rho_\mu(\chi)}{\partial\chi} = n\left(\hat{\rho}_\mu(\chi)\right)^\frac{n-1}{n} \left((\rho_\mu^+)^\frac{1}{n} - (\rho_\mu^-)^\frac{1}{n}\right)
\end{align}

\section{Transmission loss computation for a dynamic system}
\label{app_RT_system}

Let us consider a 2D section of a flat panel with a certain material distribution inside such that a FE discretization of the dynamic system yields
\begin{equation}
\label{eq_dynamic_system}
	\mathbb{M} \ddot{\hat{\bs{u}}} + \mathbb{C} \dot{\hat{\bs{u}}} + \mathbb{K} \hat{\bs{u}} = \hat{\bs{f}}.
\end{equation} 
The domain is assumed infinite in the vertical direction and in contact with air at both sides in the horizontal direction. A plane wave travels at a certain frequency in the air domain in contact with the left side of the panel and it continues to propagate at the same frequency once it reaches the right side of the panel. Note that, since the analysis is performed for a given frequency, the system \eqref{eq_dynamic_system} may be expressed in the frequency domain as
\begin{equation}
\label{eq_dynamic_system_freq}
	\mathbb{D}(\omega) \hat{\bs{U}} = \hat{\bs{F}}, \quad \mathbb{D}(\omega) = \mathbb{K} - i\omega \mathbb{C} - \omega^2 \mathbb{M}.
\end{equation}
Since the air domain is assumed to spread infinitely at both sides, there will be two waves on the left side propagating in opposite directions perpendicular to the panel's surface as a result of the incident wave reflection, while only one wave will be transmitted to the air in the right side. The analytical solutions for the displacement and pressure fields of these waves are given by equations \eqref{eq_comp1} and \eqref{eq_comp12} for the air on the left side and by equations \eqref{eq_comp21} and \eqref{eq_comp2} for the air on the right side. Note that the reflection and transmission coefficients, $R$ and $T$ respectively, are the unknowns to be solved in this problem. 

In order to express the system \eqref{eq_dynamic_system_freq} in terms of $R$ and $T$ compatibility conditions for the horizontal component of the displacements and pressure will be applied, which yields:
\begin{align}
	&\hat{\bs{U}} =
	\begin{bmatrix}
		\hat{\bs{U}}^{(i)}\\
		\hat{\bs{U}}^{(b)}\\
		\hat{\bs{U}}^{(t)}\\
		\hat{\bs{U}}^{(l)}\\
		\hat{\bs{U}}^{(r)}
	\end{bmatrix} = 
	\underbrace{\begin{bmatrix}
		\bs{I} & \bs{0} &  \bs{0} & \bs{0} \\
		\bs{0} & \bs{I} &  \bs{0} & \bs{0} \\
		\bs{0} & \bs{I} &  \bs{0} & \bs{0} \\
		\bs{0} & \bs{0} & -\bs{1} & \bs{0} \\
		\bs{0} & \bs{0} &  \bs{0} & \bs{1} \\		
	\end{bmatrix}}_{\mathbb{P}_\bs{u}}
	\underbrace{\begin{bmatrix}
		\hat{\bs{U}}^{(i)}\\
		\hat{\bs{U}}^{(b)}\\
		R\\
		T
	\end{bmatrix}}_{\hat{\bs{U}}_1} + 
	\underbrace{\begin{bmatrix}
		\bs{0} \\
		\bs{0} \\
		\bs{0} \\
		\bs{1} \\
		\bs{0}
	\end{bmatrix}}_{\hat{\bs{U}}_0} \\
	&\hat{\bs{F}} =
	\begin{bmatrix}
	\hat{\bs{F}}^{(i)}\\
	\hat{\bs{F}}^{(b)}\\
	\hat{\bs{F}}^{(t)}\\
	\hat{\bs{F}}^{(l)}\\
	\hat{\bs{F}}^{(r)}
	\end{bmatrix} = -i K_a 
	\underbrace{\left(\begin{bmatrix}
	\bs{0} & \bs{0} &  \bs{0} & \bs{0} \\
	\bs{0} & \bs{0} &  \bs{0} & \bs{0} \\
	\bs{0} & \bs{0} &  \bs{0} & \bs{0} \\
	\bs{0} & \bs{0} &  \bs{1} & \bs{0} \\
	\bs{0} & \bs{0} &  \bs{0} & \bs{1} \\		
	\end{bmatrix}\right.}_{\mathbb{P}_\bs{f}}
	\underbrace{\begin{bmatrix}
	\hat{\bs{U}}^{(i)}\\
	\hat{\bs{U}}^{(b)}\\
	R\\
	T
	\end{bmatrix}}_{\hat{\bs{U}}_1} + 
	\underbrace{\left.\begin{bmatrix}
	\bs{0} \\
	\bs{0} \\
	\bs{0} \\
	\bs{1} \\
	\bs{0}
	\end{bmatrix}\right)}_{\hat{\bs{U}}_0},
\end{align}
where $\bs{I}$ are identity matrices, $\bs{0}$ refer to matrices/vectors of zeros, $\bs{1}$ are column vectors of ones, $K_a=\rho_a v_a \omega S$ (with $S$ being the panel's surface area in contact with the air at each side) and the superscripts $(l)$ and $(r)$ refer to the left and right side horizontal degrees of freedom, respectively, $(b)$ and $(t)$ refer to both horizontal and vertical degrees of freedom at the bottom and top sides of the domain, respectively, and $(i)$ refers to the remaining degrees of freedom. Pre-multiplying the system \eqref{eq_dynamic_system_freq} by $\mathbb{P}_\bs{u}^\text{T}$ yields
\begin{equation}
	\underbrace{(\mathbb{P}_\bs{u}^\text{T} \mathbb{D} \mathbb{P}_\bs{u} + iK_a \mathbb{P}_\bs{u}^\text{T}\mathbb{P}_\bs{f})}_\mathbb{A} \hat{\bs{U}}_1 = \underbrace{-\mathbb{P}_\bs{u}^\text{T}(\mathbb{D} + i K_a \bs{I}) \hat{\bs{U}}_0 }_{\hat{\bs{B}}}.
\end{equation}
Now, defining
\begin{equation}
	\hat{\bs{U}}^{(f)} =
	\begin{bmatrix}
	\hat{\bs{U}}^{(i)} \\
	\hat{\bs{U}}^{(b)} \\
	\end{bmatrix}
\end{equation}
allows one to express
\begin{equation}
\label{eq_system_ext}
\begin{bmatrix}
	\mathbb{A}^{(ff)} & \bs{A}^{(fL)} & \bs{A}^{(fR)} \\
	\bs{A}^{(Lf)} & A^{(LL)} & A^{(LR)} \\
	\bs{A}^{(Rf)} & A^{(RL)} & A^{(RR)}
\end{bmatrix}
\begin{bmatrix}
	\hat{\bs{U}}^{(f)} \\
	R \\
	T 
\end{bmatrix} = 
\begin{bmatrix}
\hat{\bs{B}}^{(f)} \\
\hat{B}^{(L)} \\
\hat{B}^{(R)} 
\end{bmatrix}.
\end{equation}
The system in equation \eqref{eq_system_ext} can be reduced by first expressing
\begin{equation}
\label{eq_internal_disp}
	\hat{\bs{U}}^{(f)} = (\mathbb{A}^{(ff)})^{-1}(\hat{\bs{B}}^{(f)}-\bs{A}^{(fL)}R-\bs{A}^{(fR)}T)
\end{equation}
and then introducing expression \eqref{eq_internal_disp} into the second and third equations of system \eqref{eq_system_ext}, so that
\begin{equation}
\label{eq_system_red}
	\begin{bmatrix}
	\bar{A}^{(LL)} & \bar{A}^{(LR)} \\
	\bar{A}^{(RL)} & \bar{A}^{(RR)}
	\end{bmatrix}
	\begin{bmatrix}
	R \\
	T
	\end{bmatrix} = 
	\begin{bmatrix}
	\bar{B}^{(L)} \\
	\bar{B}^{(R)}
	\end{bmatrix},
\end{equation}
where, for each row $a=\{L,R\}$ and column $b=\{L,R\}$,
\begin{align}
	&\bar{A}^{(ab)} = A^{(ab)} - \bs{A}^{(af)}(\mathbb{A}^{(ff)})^{-1}\bs{A}^{(fb)}, \\
	&\bar{B}^{(a)} = \hat{B}^{(a)} - \bs{A}^{(af)}(\mathbb{A}^{(ff)})^{-1}\hat{\bs{B}}^{(f)}.
\end{align}
Note that the system in equation \eqref{eq_system_red} is a complex $2\times2$ system, the solution of which gives the reflection and transmission coefficients:
\begin{align}
	R = \dfrac{ \bar{A}^{(RR)}\bar{B}^{(L)}-\bar{A}^{(LR)}\bar{B}^{(R)} }{ \bar{A}^{(LL)}\bar{A}^{(RR)}-\bar{A}^{(RL)}\bar{A}^{(LR)} }, \\
	T = \dfrac{ \bar{A}^{(LL)}\bar{B}^{(R)}-\bar{A}^{(RL)}\bar{B}^{(L)} }{ \bar{A}^{(LL)}\bar{A}^{(RR)}-\bar{A}^{(RL)}\bar{A}^{(LR)} }.
\end{align}

The transmission loss is finally obtained by
\begin{equation}
	\text{TL} = -20 \log_{10} |T|,
\end{equation}
where $|T|$ refers to the complex module of the transmission coefficient $T$.

\section*{References}
\bibliographystyle{plainnat}
\bibliography{Biblio}
	
\end{document}